\documentstyle[prl,eqsecnum,aps]{revtex}

\begin{document}
\author{Jian Qi Shen\footnote{Electronic address: jqshen@coer.zju.edu.cn, jqshencn@yahoo.com.cn}}
\address{Zhejiang Institute of Modern Physics and Department of Physics,
Zhejiang University {\rm (}Yuquan Campus{\rm )}, \\Hangzhou
310027, People's republic of China}
\date{\today }
\title{Dual curvature tensors and dynamics of gravitomagnetic matter}
\maketitle

\begin{abstract}
Gravitomagnetic charge that can also be referred to as the {\it
dual mass} or {\it magnetic mass} is the topological charge in
gravity theory. A gravitomagnetic monopole at rest can produce a
stationary gravitomagnetic field. Due to the topological nature of
gravitomagnetic charge, the metric of spacetime where the
gravitomagnetic matter is present will be nonanalytic. In this
paper both the dual curvature tensors (which can characterize the
dynamics of gravitational charge/monopoles) and the antisymmetric
gravitational field equation of gravitomagnetic matter are
presented. We consider and discuss the mathematical formulation
and physical properties of the dual curvature tensors and scalar,
antisymmetric source tensors, dual spin connection (including the
low-motion weak-field approximation), dual vierbein field as well
as dual current densities of gravitomagnetic charge. It is shown
that the dynamics of gravitomagnetic charge can be founded within
the framework of the above dual quantities. In addition, the dual
relationship between the dynamical theories of gravitomagnetic
charge (dual mass) and gravitoelectric charge (mass) is also taken
into account in the present paper.
\\ \\


\end{abstract}

\section{Introduction}
The gravitomagnetic monopole possesses the dual charge to the
mass. In this sense, the gravitomagnetic charge can also be
referred to as the {\it dual mass}\cite{Shen,Shenarxiv}. Some
authors referred to it as ``magnetic mass'' or {\it magnetic-type
mass} ({\it magnetic-like mass})\cite{Zimmerman}. Historically,
Newman {\it et al.} discovered a stationary and spherically
symmetric solution (now known as the NUT solution) that contains a
second parameter $\iota$ besides the mass $m$, and regarded it as
the empty-space generalization of the Schwarzschild
solution\cite{Newman}. Demiansky and Newman found that the NUT
space is in fact the spacetime produced by a mass which has a
gravitomagnetic charge\cite{Demiansky}. Dowker and Roche
independently rediscovered this interpretation of the NUT
parameter $\iota$\cite{Dowker}. The cylindrically symmetric
solution of gravitomagnetic charge ({\it i.e.}, the cylindrical
analogue of NUT space, which is the spacetime of a line
gravitomagnetic monopole)\cite{Nouri-Zonoz} was considered by
Nouri-Zonoz in 1997. The various properties of NUT space attracts
attention of some authors\cite{Zimmerman,Lynden-Bell,Miller}.
Lynden-Bell and Nouri-Zonoz gave a new and more elementary
derivation of NUT space based on the spherical symmetry of
gravitoelectric and gravitomagnetic fields of NUT space and the
spherical symmetry of its spatial metric
$\gamma_{ab}$\cite{Lynden-Bell}. They proved that all geodesics of
NUT space lie on spatial cones, and used this interesting theorem
to determine the gravitational lensing properties of NUT
space\cite{Lynden-Bell}. Miller studied the global properties of
the Kerr-Taub-NUT metric\cite{Miller}. Zimmerman and Shahir
considered the geodesics in the NUT metric and found that the NUT
geodesics are similar to the properties of trajectories for
charged particles orbiting about a magnetic
monopole\cite{Zimmerman}. Recently, the quantal field and
differential geometric properties and related topics associated
with the NUT space receive attention of several researchers. Bini
{\it et al.} investigated a single master equation for the gauge-
and tetrad- invariant first-order massless perturbation describing
spin $\leq 2$ fields in the Kerr-Taub-NUT spacetime\cite{Bini1}.
In their another paper, Bini {\it et al.} analyzed the parallel
transport around closed circular orbits in the equatorial plane of
the Taub-NUT spacetime to reveal the effect of the gravitomagnetic
monopole parameter on circular holonomy
transformations\cite{Bini2}. As far as the quantization of
gravitomagnetic monopole is concerned, historically, many authors
considered this problem by analogy with Dirac's argument for the
quantization of magnetic charge\cite{Dowker,Dowker2,Zee}. In
history, it was clear to see that such an idea (as Dirac's theory)
might not lead to a consistent quantum field theory for the
gravitomagnetic charge\cite{Lynden-Bell} if the gravitomagnetic
charge is truly present in the universe. This topic will be
further taken into consideration in our other work regarding the
quantum theory of gravitomagnetic monopole.

Lynden-Bell and Nouri-Zonoz thought that it is natural to ask how
general relativity must be modified to allow for gravitomagnetic
monopole densities and currents\cite{Lynden-Bell}. They
conjectured the generalization will be to spaces with
unsymmetrical affine connections which have nonzero
torsion\cite{Lynden-Bell}. However, by analogy with the
electrodynamics which is only changed to include
$\partial_{\mu}\tilde{{\mathcal F}}^{\mu\nu}=j^{\nu}_{\rm m}$ but
does not come with an additional vector
potential\cite{Lynden-Bell,Zeleny} if the magnetic charge truly
exists, here we will suggest another alternative to this problem,
{\it i.e.}, constructing an antisymmetric dual Einstein tensor to
describe the gravitational field produced by gravitomagnetic
matter. Note that here we will not introduce extra unsymmetrical
affine connections.

This paper has a threefold purpose: to establish the theoretical
basis (such as dual curvature tensors) on which the field equation
of gravitomagnetic matter is suggested; to apply the gravitational
field equation to some problems (such as the NUT metric); to
consider other related topics (such as the dual metric, dual spin
connection and dual current density). The physical reasons for the
nature of the mathematical results obtained will be interpreted
and discussed wherever possible. The paper presents a detailed and
self-contained treatment of the fundamental problems of classical
gravitomagnetic charge.

The paper is organized as follows: in Sec. II, by considering the
connection among the NUT solution, Einstein's equation and the
dynamics of gravitomagnetic matter, it is shown that
gravitomagnetic matter requires a gravitational field equation of
its own; in Sec. III, we study the dual Riemann curvature tensors
and dual Ricci curvature tensors. The relationship between these
dual curvature tensors and the Einstein (and Ricci) tensor is also
demonstrated. In Sec. IV, we take into consideration the
comparison of the gravity theory (involving the gravitomagnetic
charge) with the electrodynamics, which will provide clue to us on
how to construct the gravitational field equation of dual mass
(and hence its dynamics). Several fundamental subjects regarding
the dynamics of gravitomagnetic matter are discussed in Sec. V,
which are as follows: (i) by using the variational principle, the
antisymmetric dual Einstein tensor is obtained with the dual
curvature scalar; (ii) the antisymmetric source tensor of dual
matter is constructed in terms of the four-dimensional velocity
and the corresponding covariant derivatives; (iii) the dual spin
connection which will arise in the field equation of dual mass is
briefly suggested. The dual vierbein field, dual metric and dual
current density of gravitomagnetic matter are taken into account
in Sec. VI. The importance of these quantities derives from the
fact that many dynamical variables (such as the source tensors) of
physical interest regarding the gravitomagnetic matter can be
expressed in terms of them. To consider the connection between the
gravitomagnetic charge and the magnetic charge, in Sec. VII, we
study the five-dimensional field equation of gravitomagnetic
charge. In Sec. VIII, we discuss briefly the duality relationship
between the dynamical theories of gravitomagnetic charge (dual
mass) and gravitoelectric charge (mass). In Sec. IX, we conclude
with some remarks.

\section{Wanting a gravitational field equation necessary}
Is gravitomagnetic monopoles required of its own gravitational
field equation for dealing with the gravitational effects
(properties, phenomena)?  Before
 proceeding with the treatment of dynamics of gravitomagnetic
 matter, we shall here consider a problem associated with the
 relationship among the field equation, the solutions and the Bianchi
 identities. In an attempt to investigate the metric resulting from the presence of the gravitomagnetic
charge currents, we think that one may meet with problems
concerning the choice of the ``true'' metric produced by
gravitomagnetic charge from the nonanalytic solutions of
Einstein's field equation. We argue that in history the
gravitational field equation of gravitomagnetic charge might get
less attention than it deserves.

 As is known to us all, in general relativity, the NUT metric (named after Newman, Tamborino and
 Unti) $g_{\mu\nu}(m,\iota)$ is the one describing the
 gravitational distribution produced by both the gravitoelectric charge (with mass $m$) and
 the gravitomagnetic monopole (with dual mass\footnote{Dual mass is viewed as the gravitomagnetic monopole strength.} $\iota$) fixed at the origin of the
 coordinate system\cite{Demiansky}. Since it is well known that the NUT solution $g_{\mu\nu}(m=0,\iota\neq 0)$
 satisfies Einstein's vacuum equation ${\mathcal R}_{\mu\nu}=0$ (and hence Einstein's equation $G_{\mu\nu}=0$ without
 matter), someone may hold that the dynamics of gravitomagnetic matter has already been embodied in
 Einstein'gravitational field equation, or that the gravitational field equation of
 gravitomagnetic charge has been involved in Einstein's equation,
 or that the gravitational distribution of both mass and dual mass
 is governed together by Einstein's equation. Thus from the point of view of these peoples it is concluded that
 there exists no other gravitational field equation of gravitomagnetic matter than Einstein's equation,
 or that the potential suggestion of
 its gravitational field equation is not
 essential in investigating the dynamics of gravitomagnetic
 monopole (should such exist). However, even though it sounds
 reasonable, we think that these viewpoints may be not the true case.
 We believe that, in fact, the gravitomagnetic matter
 needs its own gravitational field equation rather
than Einstein's equation, since the latter governs the spacetime
of gravitoelectric matter (mass) only. In what follows it will
just be concentrated on the field equation problem of
gravitomagnetic charge by comparing it with the electromagnetic
field equation of magnetic monopole.

The NUT solution $g_{\mu\nu}(m=0,\iota\neq 0)$, which are the
nonanalytic functions, truly agrees with Einstein's equation
(empty-spacetime field equation), and it can truly describe the
gravitational field near a gravitomagnetic monopole. This,
however, does not means that all the {\it nonanalytic} solutions
to Einstein's equation belong to the contribution of
gravitomagnetic charge. Namely, Einstein's equation is
overdetermined for determining the gravitational-potential
solutions of gravitomagnetic charge. Moreover, it can be further
concluded that the NUT solution $g_{\mu\nu}(m=0,\iota\neq 0)$
satisfying Einstein's equation is not necessarily the one only to
Einstein's equation, just like the case that a solution satisfying
a Bianchi identity is not the one only to this Bianchi identity.

In order to clarify the above viewpoint further, let us consider
an illustrative example. Since there are some analogies between
general relativity and electromagnetism, and most of the essential
features of dynamics of electric and magnetic charges share
similarity to the gravity theories of gravitoelectric and
gravitomagnetic charges, one can prove the previous interpretation
to be sound by analogy with the Maxwellian equation in
electrodynamics. It is well known that the electromagnetic field
equation of electric charge and magnetic charge are written in the
form $\partial_{\mu}{\mathcal F}^{\mu\nu}=j^{\nu}_{\rm e}$ and
$\partial_{\mu}\tilde{{\mathcal F}}^{\mu\nu}=j^{\nu}_{\rm m}$,
respectively. Here, the dual electromagnetic field tensor
$\tilde{{\mathcal
F}}^{\mu\nu}=\frac{1}{2}\epsilon^{\mu\nu\alpha\beta}{\mathcal
F}_{\alpha\beta}$ with $\epsilon^{\mu\nu\alpha\beta}$ being the
Levi-Civita tensor\footnote{Note that, in this paper, for the
electrodynamics, the Levi-Civita tensor refers to that in
Minkowski flat spacetime, while for the gravity theory,
$\epsilon^{\mu\nu\alpha\beta}$ stands for the Levi-Civita tensor
in the curved spacetime.}. It is readily verified that the
stationary monopole solution (say, the Wu-Yang
solution\cite{Wu}\footnote{The Wu-Yang solution is of the form (in
spherical coordinate system) $A_{r}=0$, $A_{\theta}=0$,
$A_{\varphi}=\frac{g(1\pm\cos \theta)}{r\sin \theta}$, which is a
nonanalytic function.}) also agrees with the field equation of
electric charge ($\partial_{\mu}{\mathcal F}^{\mu\nu}=0$ without
the electric current densities). But this does not means that the
Wu-Yang solution should necessarily the one only to
$\partial_{\mu}{\mathcal F}^{\mu\nu}=0$, since the equation
($\partial_{\mu}{\mathcal F}^{\mu\nu}=0$) is overdetermined for
determining the monopole solution (for instance, it cannot set the
magnetic charge $g$ in the Wu-Yang solution). Likewise, it is
apparently seen that the electric charge solution of
$\partial_{\mu}{\mathcal F}^{\mu\nu}=j^{\nu}_{\rm e}$
automatically agrees with the electromagnetic field equation of
magnetic charge ({\it i.e.}, $\partial_{\mu}\tilde{{\mathcal
F}}^{\mu\nu}=0$ without magnetic charge densities and currents).
But it is clearly not suitable to say that the equation
$\partial_{\mu}\tilde{{\mathcal F}}^{\mu\nu}=0$ is also the
electromagnetic field equation of electric charge current
$j^{\nu}_{\rm e}$. In fact, not {\it all} the analytical solutions
which agree with $\partial_{\mu}\tilde{{\mathcal F}}^{\mu\nu}=0$
satisfy the equation $\partial_{\mu}{\mathcal
F}^{\mu\nu}=j^{\nu}_{\rm e}$. The reason for this may be as
follows: for the latter field equation, the former equation is
merely a Bianchi identity. Similarly, for the magnetic charge,
$\partial_{\mu}{\mathcal F}^{\mu\nu}=0$ should also be viewed as a
Bianchi identity, which may also be seen as follows: if the dual
electromagnetic field tensor is defined to be $\tilde{{\mathcal
F}}^{\mu\nu}=\partial^{\mu}\tilde{A}^{\nu}-\partial^{\nu}\tilde{A}^{\mu}$,
and the electromagnetic field tensor ${\mathcal F}^{\mu\nu}$
expressed in terms of the dual electromagnetic field tensor takes
the form ${\mathcal
F}^{\mu\nu}=-\frac{1}{2}\epsilon^{\mu\nu\alpha\beta}\tilde{{\mathcal
F}}_{\alpha\beta}$, then the electromagnetic field equation of
electric charge ({\it i.e.}, $\partial_{\mu}{\mathcal
F}^{\mu\nu}=0$ without the electric current densities) can just be
rewritten as the form of a Bianchi identity. Thus, we show that
the equation $\partial_{\mu}{\mathcal F}^{\mu\nu}=0$ in the
framework of dual 4-D potential vector $\tilde{A}^{\nu}$ is truly
a Bianchi identity. Since $\partial_{\mu}{\mathcal F}^{\mu\nu}=0$
is only a Bianchi identity for the magnetic charge, it cannot be
the electromagnetic field equation of magnetic monopole
(similarly, Einstein's equation in empty spacetime is not the
gravitational field equation of gravitomagnetic matter, either).
This, therefore, means that not all of the solutions satisfying
this Bianchi identity may also satisfy
$\partial_{\mu}\tilde{{\mathcal F}}^{\mu\nu}=j^{\nu}_{\rm m}$,
{\it i.e.}, not all these solutions may surely belong to the
effect of magnetic monopole (similarly, not all the empty-space
generalizations of the solutions of Einstein's equation belong to
gravitomagnetic charge, {\it i.e.}, Einstein's field equation
cannot govern independently the gravitational distribution
produced by gravitomagnetic matter). In a word, gravitomagnetic
matter should have its own gravitational field equation.

In view of the importance of gravitational field equation of
gravitomagnetic monopoles, it is certainly desirable to construct
a theoretical framework, within which the field equation (and
hence its dynamics) can be established. First and foremost, we
should study the dual curvature tensors in Riemann space.

\section{Dual curvature tensors in Riemann space}
In Sec. II, we showed that the investigation of gravitational
distribution in the vicinity of a gravitomagnetic charge/monopole
should start with the needs of a gravitational field equation. In
order to discover the gravitational field equation of
gravitomagnetic matter, we will proceed in a manner entirely
analogous to what has been done for that of magnetic monopole
({\it i.e.}, first we define the dual electromagnetic field tensor
$\tilde{{\mathcal F}}^{\mu\nu}$, and then we obtain the quantity
$\partial_{\mu}\tilde{{\mathcal F}}^{\mu\nu}$ from the variational
principle with the dual Lagrangian density).

To lay a mathematical foundation for the field equation of dual
mass that we will develop in the next section, first we consider
the properties of dual curvature tensors in Riemann space. The
right- and left- dual Riemann curvature tensors may be defined as
follows
\begin{equation}
{\mathcal
R}^{\ast}_{\mu\tau\omega\nu}=\frac{1}{2}\epsilon_{\omega\nu}^{\ \
\  \lambda\sigma}{\mathcal R}_{\mu\tau\lambda\sigma}      \quad
{\rm (right \ dual)},     \qquad   \quad      ^{\ast}{\mathcal
R}_{\mu\tau\omega\nu}=\frac{1}{2}\epsilon_{\mu\tau}^{\ \ \
\lambda\sigma}{\mathcal R}_{\lambda\sigma\omega\nu}      \quad
{\rm (left \ dual)},
\end{equation}
and consequently, the corresponding dual Ricci tensors take the
form

\begin{equation}
{\mathcal R}^{\ast}_{\mu\nu}=g^{\tau\omega}{\mathcal
R}^{\ast}_{\mu\tau\omega\nu}=-\frac{1}{2}\epsilon_{\nu}^{\ \
\tau\lambda\sigma}{\mathcal R}_{\mu\tau\lambda\sigma}      \quad
{\rm (right \ dual)},
     \qquad  \quad      ^{\ast}{\mathcal
R}_{\mu\nu}=g^{\tau\omega \ast}{\mathcal
R}_{\mu\tau\omega\nu}=\frac{1}{2}\epsilon_{\mu}^{\ \
\lambda\sigma\omega}{\mathcal R}_{\lambda\sigma\omega\nu}  \quad
{\rm (left \ dual)},
\end{equation}
respectively. Note that $\epsilon_{\nu}^{\ \
\tau\lambda\sigma}{\mathcal R}_{\mu\tau\lambda\sigma}\equiv 0$,
{\it i.e.}, the right-dual Ricci tensor ${\mathcal
R}^{\ast}_{\mu\nu}$ is identically zero.  So, we think that
${\mathcal R}^{\ast}_{\mu\nu}$ may have no important value in
dealing with the dynamics of dual mass and therefore we abandon
it\footnote{It should be pointed out that in the published paper
[Gen. Relativ. Gravit. {\bf 34}, 1423-1435 (2002)], however, we
adopted the right-dual form rather than the left-dual one. Such a
situation was caused by the fact that the form of the Riemann
curvature tensor ${\mathcal R}_{\mu\nu\alpha\beta}$ applied in the
GRG paper (see, for example, Eq.(7) in the above-mentioned GRG
paper) refers to the form ${\mathcal R}_{\alpha\beta\mu\nu}$
expressed in the present paper.} and only adopt the left-dual
curvature tensors in the following discussion. In particular, in
what follows, the left-dual Riemann tensor and its corresponding
dual Ricci tensor will be denoted respectively by
$\tilde{{\mathcal R}}_{\mu\tau\omega\nu}$ and $\tilde{{\mathcal
R}}_{\mu\nu}$, {\it i.e.},
\begin{equation}
\tilde{{\mathcal R}}_{\mu\tau\omega\nu}\equiv\ ^{\ast}{\mathcal
R}_{\mu\tau\omega\nu}=\frac{1}{2}\epsilon_{\mu\tau}^{\ \ \
\lambda\sigma}{\mathcal R}_{\lambda\sigma\omega\nu},       \qquad
\tilde{{\mathcal R}}_{\mu\nu}\equiv \ ^{\ast}{\mathcal
R}_{\mu\nu}=\frac{1}{2}\epsilon_{\mu}^{\ \
\lambda\sigma\omega}{\mathcal R}_{\lambda\sigma\omega\nu}.
\label{eq33}
\end{equation}
It should be pointed out that in this section, without loss of
generality, we shall restrict ourselves to the left- and
right-dual curvature tensors only. For other dual curvature
tensors (such as mid-dual and two-side dual tensors) one may be
referred to the appendix to this paper.

To consider further the properties of dual curvature tensor, we
will present the following three interesting and useful identities
regarding the above-mentioned dual curvature tensors:

\begin{eqnarray}
& &  \frac{1}{2}\epsilon^{\theta\tau\omega\nu}\tilde{{\mathcal R}}^{\mu}_{\ \ \tau\omega\nu}=-\frac{1}{2}\left(G^{\theta\mu}+G'^{\theta\mu}\right),                    \nonumber \\
& &  \frac{1}{2}\epsilon^{\theta\tau\omega\nu}\tilde{{\mathcal
R}}_{\tau\omega\nu}^{\ \ \ \ \mu}=-\frac{1}{2}\left({\mathcal
R'}^{\theta\mu}+{\mathcal R}^{\theta\mu}\right),
                 \nonumber \\
& &  \frac{1}{2}\epsilon^{\mu\nu\alpha\beta}\left(\tilde{{\mathcal
R}}_{\mu\nu}-\tilde{{\mathcal
R}}_{\nu\mu}\right)=-\frac{1}{2}\left[\left({\mathcal
R}^{\alpha\beta}-{\mathcal R}^{\beta\alpha}\right)+\left({\mathcal
R'}^{\alpha\beta}-{\mathcal R'}^{\beta\alpha}\right)\right],
\nonumber \\
& & \frac{1}{2}\epsilon^{\mu\nu\alpha\beta}\cdot \frac{1}{2}
\left[\left({\mathcal R}_{\mu\nu}-{\mathcal
R}_{\nu\mu}\right)+\left({\mathcal R'}_{\mu\nu}-{\mathcal
R'}_{\nu\mu}\right)\right]=\tilde{{\mathcal
R}}^{\alpha\beta}-\tilde{{\mathcal R}}^{\beta\alpha}. \label{eq34}
\end{eqnarray}
Here ${\mathcal R'}^{\alpha\beta}={\mathcal R}^{\alpha\lambda}_{\
\ \ \lambda}\ ^{\beta}$, ${\mathcal R'}_{\mu\nu}={\mathcal
R}_{\mu}^{\ \ \lambda}\ _{\lambda\nu}$. The proof of the first
identity in Eq.(\ref{eq34}) is given in Appendix 2 to this paper.
Now let us look at the second identity in Eq.(\ref{eq34}), which
can be easily verified as follows:
\begin{equation}
\frac{1}{2}\epsilon^{\theta\tau\omega\nu}\tilde{{\mathcal
R}}_{\tau\omega\nu}^{\ \ \ \
 \mu}=\frac{1}{4}\epsilon^{\theta\tau\omega\nu}\epsilon_{\tau\omega}^{\
\ \ \lambda\sigma}{\mathcal R}_{\lambda\sigma\nu}^{\ \ \ \
\mu}=-\frac{1}{2}\left(g^{\lambda\theta}g^{\sigma\nu}-g^{\lambda\nu}g^{\sigma\theta}\right){\mathcal
R}_{\lambda\sigma\nu}^{\ \ \ \ \mu}=-\frac{1}{2}\left({\mathcal
R}^{\theta\nu}_{\ \ \ \nu}\ ^{\mu}-{\mathcal R}^{\nu\theta}_{\ \ \
\nu}\ ^{\mu}\right),
\end{equation}
where for the calculation of
$\epsilon^{\theta\tau\omega\nu}\epsilon_{\tau\omega}^{\ \ \
\lambda\sigma}$, readers may be referred to Appendix 1. In
Appendix 1 the Ricci tensor is so defined that ${\mathcal
R}^{\nu\theta\mu}_{\ \ \ \ \nu}={\mathcal R}^{\theta\mu}$. By
using ${\mathcal R}^{\theta\nu}_{\ \ \ \nu}\ ^{\mu}={\mathcal
R'}^{\theta\mu}$, the second identity in Eq.(\ref{eq34}) is proven
correct. With the help of the formulae\footnote{It follows from
the formula presented in Appendix 1 to this paper that the inner
product of two Levi-Civita tensors/symbols (where the summation is
to be carried out over the repeated index $\mu=0,1,2,3$) reads
$\epsilon^{\mu\nu\alpha\beta}\epsilon_{\mu}^{\ \
\lambda\sigma\omega}=-\left(g^{\lambda\nu}g^{\sigma\alpha}g^{\omega\beta}+
g^{\lambda\beta}g^{\sigma\nu}g^{\omega\alpha}+g^{\lambda\alpha}g^{\sigma\beta}
g^{\omega\nu}-g^{\lambda\beta}g^{\sigma\alpha}g^{\omega\nu}-g^{\lambda\nu}
g^{\sigma\beta}g^{\omega\alpha}-g^{\lambda\alpha}g^{\sigma\nu}g^{\omega\beta}\right)$.}
in Appendix 1, one can arrive at

\begin{eqnarray}
\frac{1}{2}\epsilon^{\mu\nu\alpha\beta}\tilde{{\mathcal
R}}_{\mu\nu}&=&\frac{1}{4}\epsilon^{\mu\nu\alpha\beta}\epsilon_{\mu}^{\
\ \lambda\sigma\omega}{\mathcal R}_{\lambda\sigma\omega\nu}                 \nonumber \\
&=&-\frac{1}{4}\left({\mathcal R}^{\lambda\alpha\beta}_{\ \ \ \
\lambda }+{\mathcal R}^{\beta\nu\alpha}_{\ \ \ \ \nu}+{\mathcal
R}^{\alpha\beta\nu}_{\ \ \ \ \nu}-{\mathcal R}^{\beta\alpha\nu}_{\
\ \ \ \nu }-{\mathcal R}^{\nu\beta\alpha}_{\ \ \ \ \nu}-{\mathcal
R}^{\alpha\nu\beta}_{\ \ \ \
\nu}\right)=-\frac{1}{4}\left({\mathcal R}^{\alpha\beta}-{\mathcal
R}^{\beta\alpha}\right)-\frac{1}{4}\left({\mathcal
R'}^{\alpha\beta}-{\mathcal R'}^{\beta\alpha}\right).
\end{eqnarray}
Thus we obtain
\begin{equation}
\frac{1}{2}\epsilon^{\mu\nu\alpha\beta}\tilde{{\mathcal
R}}_{\mu\nu}=-\frac{1}{4}\left[\left({\mathcal
R}^{\alpha\beta}-{\mathcal R}^{\beta\alpha}\right)+\left({\mathcal
R'}^{\alpha\beta}-{\mathcal R'}^{\beta\alpha}\right)\right],
\qquad -\frac{1}{2}\epsilon^{\mu\nu\alpha\beta}\tilde{{\mathcal
R}}_{\nu\mu}=-\frac{1}{4}\left[\left({\mathcal
R}^{\alpha\beta}-{\mathcal R}^{\beta\alpha}\right)+\left({\mathcal
R'}^{\alpha\beta}-{\mathcal R'}^{\beta\alpha}\right)\right].
\end{equation}
It follows that
\begin{equation}
\frac{1}{2}\epsilon^{\mu\nu\alpha\beta}\left(\tilde{{\mathcal
R}}_{\mu\nu}-\tilde{{\mathcal
R}}_{\nu\mu}\right)=-\frac{1}{2}\left[\left({\mathcal
R}^{\alpha\beta}-{\mathcal R}^{\beta\alpha}\right)+\left({\mathcal
R'}^{\alpha\beta}-{\mathcal R'}^{\beta\alpha}\right)\right],
\end{equation}
which is the third identity in Eq.(\ref{eq34}). Furthermore, by
the aid of the formula in Appendix 1, one can show that the fourth
identity in Eq.(\ref{eq34}) holds also.

In the following we will briefly discuss the difference between
${\mathcal R}_{\mu\nu}$ and ${\mathcal R'}_{\mu\nu}$. For the
Riemann tensor, even though ${\mathcal
R}_{\alpha\mu\nu\beta}=-{\mathcal R}_{\alpha\mu\beta\nu}$ holds,
the relation ${\mathcal R}_{\alpha\mu\nu\beta}=-{\mathcal
R}_{\mu\alpha\nu\beta}$ is no longer valid due to the nonanalytic
property of the metric tensors, which can be seen from the
following calculation:
\begin{equation}
{\mathcal R}_{\alpha\mu\nu\beta}+{\mathcal
R}_{\mu\alpha\nu\beta}=\partial_{\nu}\left(\Gamma_{\alpha,\mu\beta}+\Gamma_{\mu,\alpha\beta}\right)-\partial_{\beta}\left(\Gamma_{\alpha,\mu\nu}+\Gamma_{\mu,\alpha\nu}\right)=\left(\partial_{\nu}\partial_{\beta}-\partial_{\beta}\partial_{\nu}\right)g_{\alpha\mu}\neq
0.
\end{equation}
Thus we have ${\mathcal R}_{\alpha\mu\nu\beta}={\mathcal
R}_{\mu\alpha\beta\nu}+\left(\partial_{\nu}\partial_{\beta}-\partial_{\beta}\partial_{\nu}\right)g_{\alpha\mu}$.
With the help of ${\mathcal R}_{\mu\nu}=g^{\alpha\beta}{\mathcal
R}_{\alpha\mu\nu\beta}$ and ${\mathcal
R'}_{\mu\nu}=g^{\alpha\beta}{\mathcal R}_{\mu\alpha\beta\nu}$, one
can arrive at
\begin{equation}
{\mathcal R}_{\mu\nu}={\mathcal
R'}_{\mu\nu}+g^{\alpha\beta}\left(\partial_{\nu}\partial_{\beta}-\partial_{\beta}\partial_{\nu}\right)g_{\alpha\mu},
\end{equation}
which is the relation between ${\mathcal R}_{\mu\nu}$ and
${\mathcal R'}_{\mu\nu}$. Apparently, if $g_{\alpha\mu}$ is an
analytic function, then we have ${\mathcal R}_{\mu\nu}={\mathcal
R'}_{\mu\nu}$. It should be noted that in the dynamics of
gravitomagnetic matter, the metric is often nonanalytic. For this
reason, one should distinguish ${\mathcal R'}_{\mu\nu}$ from
${\mathcal R}_{\mu\nu}$ in the treatment of the problems where
both mass and dual mass are involved.

\section{The connection between the gravity theory and the electrodynamics}
It is useful to consider the analogies between the gravity theory
and the electrodynamics in developing the gravitational field
equation of dual matter. This deep connection between them will
make the establishment of gravity theory of gravitomagnetic matter
more enlightening and possible.

The first analogy between them is that the Lagrangian density of
electromagnetic fields can be constructed in terms of the dual
field tensor $\tilde{{\mathcal F}}_{\mu\nu}$, {\it i.e.},
\begin{equation}
-\frac{1}{4}\tilde{{\mathcal
F}}_{\mu\nu}\tilde{F}^{\mu\nu}=-\left(-\frac{1}{4}{\mathcal
F}_{\mu\nu}{\mathcal F}^{\mu\nu}\right)   \quad   {\rm with} \quad
\tilde{{\mathcal F}}_{\mu\nu}=\frac{1}{2}\epsilon_{\mu\nu}^{\ \ \
\alpha\beta }{\mathcal F}_{\alpha\beta},
\end{equation}
the gravitational analogy to which arises also. By making use of
the identities of Eq.(\ref{eq34}), it is shown that
\begin{equation}
\frac{1}{2}\epsilon^{\mu\tau\omega\nu}\tilde{{\mathcal
R}}_{\tau\omega\nu\mu}=-{\mathcal R}      \label{eq42}
\end{equation}
with ${\mathcal R}$ being the curvature scalar.

The second analogy is that one can obtain the term
$\partial_{\mu}\tilde{{\mathcal F}}^{\mu\nu}$ that appears on the
left-handed side of the electromagnetic field equation of magnetic
charge currents from the variational principle with the dual
Lagrangian density ${\mathcal L}=-\frac{1}{4}\tilde{{\mathcal
F}}_{\mu\nu}{\mathcal F}^{\mu\nu}$. Note that here the variational
principle is applied to ${\mathcal L}$ with respect to $A^{\mu}$.
In the similar manner, for the gravity theory, the dual Einstein
tensor $\tilde{G}_{\mu\nu}$ can also be derived via the
variational principle with the dual curvature scalar
$\tilde{{\mathcal R}}$, {\it i.e.},
\begin{equation}
\delta \int_{\Omega}\sqrt{-g}\tilde{{\mathcal R}}{\rm d}\Omega=0
\Rightarrow \tilde{G}_{\mu\nu}=\tilde{{\mathcal
R}}_{\mu\nu}-\tilde{{\mathcal R}}_{\nu\mu},         \label{eqG}
\end{equation}
where $\Omega $ denotes the spacetime region of volume integral
and ${\rm d}\Omega$ is the volume element. The dual curvature
scalar $\tilde{{\mathcal R}}$ is defined to be $\tilde{{\mathcal
R}}=g^{\mu\nu}\tilde{{\mathcal
R}}_{\mu\nu}=\frac{1}{2}\epsilon^{\mu\lambda\sigma\tau}{\mathcal
R}_{\lambda\sigma\tau\mu}$ with $\tilde{{\mathcal
R}}_{\mu\nu}=\frac{1}{2}\epsilon_{\mu}^{\ \
\lambda\sigma\tau}{\mathcal R}_{\lambda\sigma\tau\nu}$. The
relation (\ref{eqG}) will be developed in more details in Sec. V.

The third analogy between the gravity theory and the
electrodynamics may be seen from the following three set of
expressions. The relation between the electromagnetic field tensor
${\mathcal F}_{\mu\nu}$ and its dual $\tilde{{\mathcal
F}}_{\mu\nu}$ is
\begin{equation}
\tilde{{\mathcal F}}_{\mu\nu}=\frac{1}{2}\epsilon_{\mu\nu}^{\ \ \
\alpha\beta}{\mathcal F}_{\alpha\beta},    \qquad    {\mathcal
F}_{\mu\nu}=-\frac{1}{2}\epsilon_{\mu\nu}^{\ \ \
\alpha\beta}\tilde{{\mathcal F}}_{\alpha\beta}.
\end{equation}
In the meanwhile, there exist the similar relations between
curvature tensors and their dual. For example,
\begin{equation}
\tilde{{\mathcal
R}}_{\mu\nu\tau\omega}=\frac{1}{2}\epsilon_{\mu\nu}^{\ \ \
\alpha\beta}{\mathcal R}_{\alpha\beta\tau\omega},    \qquad
{\mathcal R}_{\mu\nu\tau\omega}=-\frac{1}{2}\epsilon_{\mu\nu}^{\ \
\ \alpha\beta}\tilde{{\mathcal R}}_{\alpha\beta\tau\omega}
\end{equation}
and
\begin{equation}
\tilde{{\mathcal R}}_{\mu\nu}=\frac{1}{2}\epsilon_{\mu}^{\ \
\lambda\sigma\tau}{\mathcal R}_{\lambda\sigma\tau\nu},   \qquad
\frac{1}{2}\left({\mathcal R}_{\mu\nu}+{\mathcal
R'}_{\mu\nu}\right)=-\frac{1}{2}\epsilon_{\mu}^{\ \
\lambda\sigma\tau}\tilde{{\mathcal R}}_{\lambda\sigma\tau\nu}.
\label{eq46}
\end{equation}
Note that the first expression of Eqs.(\ref{eq46}) is due to the
definition of the dual Ricci tensor $\tilde{{\mathcal
R}}_{\mu\nu}$ in (\ref{eq33}), and the second expression of
Eqs.(\ref{eq46}) can be derived from the second identity in
Eqs.(\ref{eq34}).

We have, therefore, investigated in detail the properties of dual
curvature tensors in Riemann space. Thus, the foundation of the
dynamics of gravitomagnetic matter is laid in the preceding
sections. Moreover, we have also discussed the problem of the
requirement for a gravitational field equation of gravitomagnetic
matter of its own in studying both the gravitational field
distribution produced by the gravitomagnetic charge and the
equation of motion of gravitomagnetic monopoles. In the following
section, we will derive the gravitational field equation of
gravitomagnetic matter from the variational principle with the
dual curvature scalar.

\section{Gravitational field equation of gravitomagnetic matter}
What in gravity theory may be caused by the existence of the
gravitomagnetic monopoles? Will it lead to unsymmetrical metric
and/or to modify Einstein's field equation? As the topological
charge in spacetime, gravitomagnetic charge will inevitably result
in the nonanalytic parts in the metric functions. In this sense,
gravitomagnetic monopole does not introduce the so-called extra
gravitational potentials. So, here we need not employ the
unsymmetrical metric in the treatment of the dynamics of
gravitomagnetic matter. In order to determine the nonanalytic
parts of the metric functions caused by the gravitomagnetic
monopoles, we should first suggest a gravitational field equation
of its own. A systematic procedure for the gravitational field
equation of gravitomagnetic matter will be developed in this
section. It will lay a physical foundation for the mathematical
formalism in Sec. IV. If the gravitomagnetic charge is not
present, then the contribution of this gravitational field
equation without source term in general relativity is easily shown
to vanish, on account of its status of Bianchi identity. But, it
will play an essential role in treating the gravitational
distribution problem of gravitomagnetic matter (should such
exist). For examples, we shall illustrate the application of this
field equation to the NUT solution and the equation of motion of
gravitomagnetic monopoles.

\subsection{Dual curvature scalar and variational principle}
With the introduction and motivation in the previous sections, we
are now in a position to find a gravitational field equation of
gravitomagnetic matter. For this aim, one should first construct a
Lagrangian density that can describe the gravitational field with
nonanalytic property. Such a Lagrangian density should be a one
that will be vanishing if the metric function is analytic. The
only one we can choose is just the dual curvature scalar
$\tilde{{\mathcal R}}$. So, the dual action $I'$ of the
gravitational field produced by the gravitomagnetic matter may be
of the form\cite{Shen,Shenarxiv}
 \begin{equation}
I'=\int_{\Omega}\sqrt{-g}\tilde{{\mathcal R}}{\rm
d}\Omega=\int_{\Omega}\sqrt{-g}g^{\mu\nu}\left[(1-\zeta)\tilde{{\mathcal
R}}_{\mu\nu}+\zeta\tilde{{\mathcal R}}_{\nu\mu}\right]{\rm
d}\Omega,                                             \label{eq51}
\end{equation}
which is a four-dimensional volume integral of the dual curvature
scalar $\tilde{{\mathcal R}}$. Note that the dual Ricci tensor
$\tilde{{\mathcal R}}_{\mu\nu}$ is asymmetric in $\mu , \nu$. So
in Eq.(\ref{eq51}), we introduce a certain parameter $\zeta$ to
rewrite the dual action $I'$. The variation of the action $I'$ is
\begin{equation}
\delta
I'=\int_{\Omega}\delta\left(\sqrt{-g}g^{\mu\nu}\right)\left[(1-\zeta)\tilde{{\mathcal
R}}_{\mu\nu}+\zeta\tilde{{\mathcal R}}_{\nu\mu}\right]{\rm
d}\Omega
+\int_{\Omega}\sqrt{-g}g^{\mu\nu}\left[(1-\zeta)\delta\tilde{{\mathcal
R}}_{\mu\nu}+\zeta\delta\tilde{{\mathcal R}}_{\nu\mu}\right]{\rm
d}\Omega,                                     \label{eq52}
\end{equation}
The first term on the right-handed side of Eq.(\ref{eq52}) can be
rewritten as
\begin{eqnarray}
\int_{\Omega}\delta\left(\sqrt{-g}g^{\mu\nu}\right)\left[(1-\zeta)\tilde{{\mathcal
R}}_{\mu\nu}+\zeta\tilde{{\mathcal R}}_{\nu\mu}\right]{\rm
d}\Omega
&=&\int_{\Omega}\sqrt{-g}\left[(1-\zeta)\left(\tilde{{\mathcal
R}}_{\mu\nu}-\frac{1}{2}g_{\mu\nu}\tilde{{\mathcal
R}}\right)+\zeta\left(\tilde{{\mathcal
R}}_{\nu\mu}-\frac{1}{2}g_{\nu\mu}\tilde{{\mathcal
R}}\right)\right]\delta g^{\mu\nu}{\rm d}\Omega    \nonumber  \\
&=&\int_{\Omega}\sqrt{-g}\left[\left(\tilde{{\mathcal
R}}_{\mu\nu}-\frac{1}{2}g_{\mu\nu}\tilde{{\mathcal
R}}\right)-\zeta \left(\tilde{{\mathcal
R}}_{\mu\nu}-\tilde{{\mathcal R}}_{\nu\mu}\right)\right]\delta
g^{\mu\nu}{\rm d}\Omega.                            \label{eq53}
\end{eqnarray}
Now in the following we will calculate the second term on the
right-handed side of Eq.(\ref{eq52}). For convenience, we set
$A=\int_{\Omega}\sqrt{-g}g^{\mu\nu}\left[(1-\zeta)\delta\tilde{{\mathcal
R}}_{\mu\nu}+\zeta\delta\tilde{{\mathcal R}}_{\nu\mu}\right]{\rm
d}\Omega$. The infinitesimal variation of the dual Ricci tensor
$\delta \tilde{{\mathcal R}}_{\mu\nu}$ is given
\begin{equation}
\delta \tilde{{\mathcal R}}_{\mu\nu}=\frac{1}{2}\delta
\epsilon_{\mu}^{\ \ \lambda\sigma\tau}{\mathcal
R}_{\lambda\sigma\tau\nu}+\frac{1}{2}\epsilon_{\mu}^{\ \
\lambda\sigma\tau}\delta{\mathcal R}_{\lambda\sigma\tau\nu}.
\end{equation}
In Appendix 3, we show that the variation
 $\int_{\Omega}\sqrt{-g}g^{\mu\nu}\frac{1}{2}\epsilon_{\mu}^{\ \ \lambda\sigma\tau}\delta{\mathcal
R}_{\lambda\sigma\tau\nu}{\rm d}\Omega$ is actually a surface
integral over the boundary of the volume, namely, it can be
transformed into a four-dimensional volume integral of a
divergence. Since the variational principle assumes that the
variation $\delta g^{\mu\nu}$ vanishes on the boundary, it follows
that $\int_{\Omega}\sqrt{-g}g^{\mu\nu}\frac{1}{2}\epsilon_{\mu}^{\
\ \lambda\sigma\tau}\delta{\mathcal R}_{\lambda\sigma\tau\nu}{\rm
d}\Omega$ has no effect on the variation of the action and
therefore gives no contribution to the derivation of the dual
Einstein tensor $\tilde{G}_{\mu\nu}$. We can, therefore, ignore
this term\footnote{Someone may argue that because of the symmetric
property of the Christoffel symbol $\Gamma^{\lambda}_{\ \ \mu\nu}$
in the indices $\mu, \nu$, it is apparent to see that the
integrand $\sqrt{-g}g^{\mu\nu}\epsilon_{\mu}^{\ \
\lambda\sigma\tau}\delta{\mathcal R}_{\lambda\sigma\tau\nu}$ is
strictly identical to zero, and it may drop out of the variation
$\delta I'$ immediately. For the dual formalism, however, this
viewpoint is not appropriate for determining the dual Einstein
tensor. To derive the dual Einstein tensor in a stringent way, if
$\int_{\Omega}\sqrt{-g}g^{\mu\nu}\epsilon_{\mu}^{\ \
\lambda\sigma\tau}\delta{\mathcal R}_{\lambda\sigma\tau\nu}{\rm
d}\Omega$ can be truly ignored, we should first show that
$\sqrt{-g}g^{\mu\nu}\epsilon_{\mu}^{\ \
\lambda\sigma\tau}\delta{\mathcal R}_{\lambda\sigma\tau\nu}$ can
be rewritten as a covariant divergence of a certain vector
indeed.}. Thus, the term $\int_{\Omega}\sqrt{-g}g^{\mu\nu}\delta
\tilde{{\mathcal R}}_{\mu\nu}{\rm d}\Omega $ in $A$ can be reduced
to
\begin{equation}
\int_{\Omega}\sqrt{-g}g^{\mu\nu}\delta \tilde{{\mathcal
R}}_{\mu\nu}{\rm
d}\Omega=\frac{1}{2}\int_{\Omega}\sqrt{-g}g^{\mu\nu}\delta
\epsilon_{\mu}^{\ \ \lambda\sigma\tau}{\mathcal
R}_{\lambda\sigma\tau\nu}{\rm d}\Omega.        \label{eq55}
\end{equation}
By using the formula in Appendix 4, Eq.(\ref{eq55}) can be
rewritten as
\begin{eqnarray}
\int_{\Omega}\sqrt{-g}g^{\mu\nu}\delta \tilde{{\mathcal
R}}_{\mu\nu}{\rm d}\Omega
&=&\frac{1}{2}\int_{\Omega}\sqrt{-g}g^{\mu\nu}\left(-g_{\mu\alpha}g_{\theta\beta}\epsilon^{\theta\lambda\sigma\tau}+\frac{1}{2}g_{\mu\theta}g_{\alpha\beta}\epsilon^{\theta\lambda\sigma\tau}\right){\mathcal
R}_{\lambda\sigma\tau\nu}\delta
g^{\alpha\beta}{\rm d}\Omega                    \nonumber \\
&=&\int_{\Omega}\sqrt{-g}\frac{1}{2}\left(-g^{\nu}_{\alpha}\epsilon_{\beta}^{\
\ \lambda\sigma\tau }{\mathcal
R}_{\lambda\sigma\tau\nu}+\frac{1}{2}g_{\alpha\beta}\epsilon^{\nu\lambda\sigma\tau}{\mathcal
R}_{\lambda\sigma\tau\nu}\right)\delta g^{\alpha\beta}{\rm
d}\Omega
 \nonumber \\
&=&-\int_{\Omega}\sqrt{-g}\left(\tilde{{\mathcal
R}}_{\nu\mu}-\frac{1}{2}g_{\nu\mu}\tilde{{\mathcal
R}}\right)\delta g^{\mu\nu}{\rm d}\Omega
  \nonumber \\
&=&-\int_{\Omega}\sqrt{-g}\left(\tilde{{\mathcal
R}}_{\mu\nu}-\frac{1}{2}g_{\mu\nu}\tilde{{\mathcal
R}}\right)\delta g^{\mu\nu}{\rm d}\Omega.             \label{eq56}
\end{eqnarray}
In the same fashion, the term
$\int_{\Omega}\sqrt{-g}g^{\mu\nu}\delta \tilde{{\mathcal
R}}_{\nu\mu}{\rm d}\Omega $ in $A$ can be rewritten in the form
\begin{equation}
\int_{\Omega}\sqrt{-g}g^{\mu\nu}\delta \tilde{{\mathcal
R}}_{\nu\mu}{\rm
d}\Omega=-\int_{\Omega}\sqrt{-g}\left(\tilde{{\mathcal
R}}_{\mu\nu}-\frac{1}{2}g_{\mu\nu}\tilde{{\mathcal
R}}\right)\delta g^{\mu\nu}{\rm d}\Omega
=-\int_{\Omega}\sqrt{-g}\left(\tilde{{\mathcal
R}}_{\nu\mu}-\frac{1}{2}g_{\nu\mu}\tilde{{\mathcal
R}}\right)\delta g^{\mu\nu}{\rm d}\Omega.           \label{eq57}
\end{equation}
By introducing two parameters $\xi$ and $\varsigma$,
Eq.(\ref{eq56}) and (\ref{eq57}) can be rewritten as
\begin{equation}
\int_{\Omega}\sqrt{-g}g^{\mu\nu}\delta \tilde{{\mathcal
R}}_{\mu\nu}{\rm
d}\Omega=-(1-\xi)\int_{\Omega}\sqrt{-g}\left(\tilde{{\mathcal
R}}_{\mu\nu}-\frac{1}{2}g_{\mu\nu}\tilde{{\mathcal
R}}\right)\delta g^{\mu\nu}{\rm d}\Omega-\xi
\int_{\Omega}\sqrt{-g}\left(\tilde{{\mathcal
R}}_{\nu\mu}-\frac{1}{2}g_{\nu\mu}\tilde{{\mathcal
R}}\right)\delta g^{\mu\nu}{\rm d}\Omega      \label{eq58}
\end{equation}
and
\begin{equation}
\int_{\Omega}\sqrt{-g}g^{\mu\nu}\delta \tilde{{\mathcal
R}}_{\nu\mu}{\rm d}\Omega=-(1-\varsigma)
\int_{\Omega}\sqrt{-g}\left(\tilde{{\mathcal
R}}_{\nu\mu}-\frac{1}{2}g_{\nu\mu}\tilde{{\mathcal
R}}\right)\delta g^{\mu\nu}{\rm d}\Omega -\varsigma
\int_{\Omega}\sqrt{-g}\left(\tilde{{\mathcal
R}}_{\mu\nu}-\frac{1}{2}g_{\mu\nu}\tilde{{\mathcal
R}}\right)\delta g^{\mu\nu}{\rm d}\Omega.         \label{eq59}
\end{equation}
Hence, by using Eq.(\ref{eq58}) and (\ref{eq59}), one can express
the second term on the right-handed side of Eq.(\ref{eq52}) ({\it
i.e.},
 $A=\int_{\Omega}\sqrt{-g}g^{\mu\nu}\left[(1-\zeta)\delta\tilde{{\mathcal
R}}_{\mu\nu}+\zeta\delta\tilde{{\mathcal R}}_{\nu\mu}\right]{\rm
d}\Omega$) as
\begin{eqnarray}
\int_{\Omega}\sqrt{-g}g^{\mu\nu}\left[(1-\zeta)\delta\tilde{{\mathcal
R}}_{\mu\nu}+\zeta\delta\tilde{{\mathcal R}}_{\nu\mu}\right]{\rm
d}\Omega &
=&[-(1-\zeta)(1-\xi)-\zeta\varsigma]\int_{\Omega}\sqrt{-g}\left(\tilde{{\mathcal
R}}_{\mu\nu}-\frac{1}{2}g_{\mu\nu}\tilde{{\mathcal
R}}\right)\delta g^{\mu\nu}{\rm d}\Omega                \nonumber  \\
&+&[-(1-\zeta)\xi-\zeta(1-\varsigma)]\int_{\Omega}\sqrt{-g}\left(\tilde{{\mathcal
R}}_{\nu\mu}-\frac{1}{2}g_{\nu\mu}\tilde{{\mathcal
R}}\right)\delta g^{\mu\nu}{\rm d}\Omega.
\end{eqnarray}
Namely, we have
\begin{equation}
A=-\int_{\Omega}\sqrt{-g}\left(\tilde{{\mathcal
R}}_{\mu\nu}-\frac{1}{2}g_{\mu\nu}\tilde{{\mathcal
R}}\right)\delta g^{\mu\nu}{\rm
d}\Omega+[\zeta+\xi-\zeta(\xi+\varsigma)]\int_{\Omega}\sqrt{-g}\left(\tilde{{\mathcal
R}}_{\mu\nu}-\tilde{{\mathcal R}}_{\nu\mu}\right)\delta
g^{\mu\nu}{\rm d}\Omega.    \label{eq511}
\end{equation}
It follows from Eq.(\ref{eq52}), (\ref{eq53}) and (\ref{eq511})
that the variation of $I'$ is finally given
\begin{equation}
\delta
I'=[\xi-\zeta(\xi+\varsigma)]\int_{\Omega}\sqrt{-g}\left(\tilde{{\mathcal
R}}_{\mu\nu}-\tilde{{\mathcal R}}_{\nu\mu}\right)\delta
g^{\mu\nu}{\rm d}\Omega.
\end{equation}
If we set $I=\frac{1}{\xi-\zeta(\xi+\varsigma)}I'$, the expression
for the variation of $I$ reads
\begin{equation}
\delta I=\int_{\Omega}\sqrt{-g}\tilde{G}_{\mu\nu}\delta
g^{\mu\nu}{\rm d}\Omega  \quad  {\rm with}     \quad
\tilde{G}_{\mu\nu}=\tilde{{\mathcal R}}_{\mu\nu}-\tilde{{\mathcal
R}}_{\nu\mu},
\end{equation}
where $\tilde{G}_{\mu\nu}$ may be referred to as the dual Einstein
tensor, which is the resulting one for the given dual action $I$.

In what follows we will simplify the form of the dual Ricci tensor
$\tilde{{\mathcal R}}_{\mu\nu}=\frac{1}{2}\epsilon_{\mu}^{\ \
\lambda\sigma\tau}{\mathcal R}_{\lambda\sigma\tau\nu}$, which
yields
\begin{equation}
\tilde{{\mathcal R}}_{\mu\nu}=\frac{1}{4}\epsilon_{\mu}^{\ \
\lambda\sigma\tau}\frac{\partial}{\partial
x^{\tau}}\left(\frac{\partial g_{\nu\lambda}}{\partial
x^{\sigma}}-\frac{\partial g_{\sigma\nu}}{\partial
x^{\lambda}}\right).
\end{equation}
Here any symmetric parts in ${\mathcal R}_{\lambda\sigma\tau\nu}$
have simply dropped out of $\tilde{{\mathcal R}}_{\mu\nu}$ because
$\epsilon_{\mu}^{\ \ \lambda\sigma\tau}$ is completely
antisymmetric in $\lambda,\sigma,\tau$. It is physically
interesting that the $(\mu 0)$ component of the dual Ricci tensor
is $\tilde{{\mathcal R}}_{\mu0}=\frac{1}{4}\epsilon_{\mu}^{\ \
\lambda\sigma\tau}\frac{\partial}{\partial
x^{\tau}}\left(\frac{\partial g_{0\lambda}}{\partial
x^{\sigma}}-\frac{\partial g_{\sigma 0}}{\partial
x^{\lambda}}\right)$, which is somewhat analogous to its
electromagnetic counterpart $\frac{\partial}{\partial
x^{\tau}}\tilde{F}^{\tau}_{\ \ \mu}=\frac{1}{2}\epsilon^{\tau}_{\
\ \mu}\ ^{\lambda\sigma}\frac{\partial}{\partial
x^{\tau}}\left(\partial_{\lambda}A_{\sigma}-\partial_{\sigma}A_{\lambda}\right)$.
This, therefore, means that $\tilde{G}_{\mu\nu}$ may truly serve
as a dual Einstein tensor and would appear on the left-handed side
of the gravitational field equation of gravitomagnetic charge.
Clearly, the formalism for the dual Einstein tensor which we have
developed here is rigorous. It may contribute to a better
understanding of the dynamics of gravitomagnetic matter.

\subsection{Antisymmetrical source tensor}
The preceding subsection shows that the dual Einstein tensor
$\tilde{G}_{\mu\nu}$ is a second-rank antisymmetric one. So, the
source tensor of gravitomagnetic matter, which appears on the
right-handed side of the gravitational field equation of
gravitomagnetic charge, should also be an antisymmetric
one\cite{Shen,Shenarxiv}. Such antisymmetric tensors which are
constructed in terms of the four-dimensional velocity vector
$U_{\nu}$ and its covariant derivatives with respect to the
spacetime coordinates may be only the following two forms
\begin{equation}
{\mathcal K}_{\mu\nu}={\rm D}_{\mu}U_{\nu}-{\rm
D}_{\nu}U_{\mu}\equiv
\partial_{\mu}U_{\nu}-\partial_{\nu}U_{\mu},   \qquad               {\mathcal
H}_{\mu\nu}=\frac{1}{2}\epsilon_{\mu\nu}^{\ \ \
\alpha\beta}{\mathcal K}_{\alpha\beta}.
\end{equation}
So, the second-rank antisymmetric source tensor of gravitomagnetic
matter can take the following form ({\it i.e.}, the linear
combination of ${\mathcal K}_{\mu\nu}$ and ${\mathcal
H}_{\mu\nu}$)
\begin{equation}
S_{\mu\nu}=\kappa_{1}{\mathcal K}_{\mu\nu}+\kappa_{2}{\mathcal
H}_{\mu\nu},
\end{equation}
where $\kappa_{1}$ and $\kappa_{2}$ denote the two combination
coefficients.

For the Fermionic field $\psi$, its four-vector velocity (or
current density) is $ U_{\mu}=e_{\mu}^{\
a}i\bar{\psi}\gamma_{a}\psi$, where $e_{\mu}^{\ a}$ denotes the
vierbein field with $\mu, a=0,1,2,3$ and $\gamma_{a}$'s stand for
the Dirac matrices in the flat spacetime. The second-rank
antisymmetric source tensor of the gravitomagnetic Fermionic field
$\psi$ can be constructed according to the above-mentioned
definition\footnote{The tensor ${\mathcal K}_{\mu\nu}$ of
Fermionic field is given as
$i\left[\left(\partial_{\mu}\bar{\psi}\right)\gamma_{\nu}\psi-\left(\partial_{\nu}\bar{\psi}\right)\gamma_{\mu}\psi\right]+i\bar{\psi}\left(\gamma_{\nu}\partial_{\mu}-\gamma_{\mu}\partial_{\nu}\right)\psi+i\bar{\psi}\left(\partial_{\mu}\gamma_{\nu}-\partial_{\nu}\gamma_{\mu}\right)\psi$.
In the papers\cite{Shen,Shenarxiv}, the third term,
$i\bar{\psi}\left(\partial_{\mu}\gamma_{\nu}-\partial_{\nu}\gamma_{\mu}\right)\psi$,
was unfortunately neglected.}. The four-dimensional velocity (or
momentum) of the Bosonic field is $
U_{\mu}=\frac{1}{2i}\left(\varphi^{\ast}\partial_{\mu}\varphi-\varphi\partial_{\mu}\varphi^{\ast}\right)
$, and the antisymmetric tensor ${\mathcal K}_{\mu\nu}$ is
therefore $ {\mathcal
K}_{\mu\nu}=\frac{1}{i}\left(\partial_{\mu}\varphi{\ast}\partial_{\nu}\varphi-\partial_{\nu}\varphi{\ast}\partial_{\mu}\varphi\right)
$, which can be used to express the antisymmetric source tensor
$S_{\mu\nu}$ of the gravitomagnetic Bosonic field.

It should be noted that the source tensor of gravitomagnetic
matter can also be expressed in terms of the dual current density
(velocity) and the dual vierbein field $\tilde{e}_{\mu}^{\ \ a}$,
which will be discussed in Sec. VI.

\subsection{Gravitational field equation of gravitomagnetic matter}
In accordance with the above discussion, we discover the
gravitational field equation of gravitomagnetic matter as follows
(covariant form)
\begin{equation}
\tilde{G}_{\mu\nu}=S_{\mu\nu}.          \label{eq5211}
\end{equation}
Note that Einstein' field equation is of symmetry in indices. In
contrast, here the gravitational field equation is an
antisymmetric one\cite{Shen,Shenarxiv}.

So far we have obtained the gravitational field equation of
gravitomagnetic charge of its own. Now let us discuss the equation
of motion of a gravitomagnetic monopole as a test particle (or a
particle system consisting of gravitomagnetic monopoles).
According to the gravitational field equation (contravariant form)
\begin{equation}
\tilde{G}^{\mu\nu}=S^{\mu\nu},                 \label{eq521}
\end{equation}
we set a second-rank antisymmetric tensor ${\mathcal
Z}^{\mu\nu}=S^{\mu\nu}-\tilde{G}^{\mu\nu}$, the covariant
divergence of which is expressed by
\begin{equation}
{\mathcal Z}^{\mu\nu}_{\ \
;\nu}=\frac{1}{\sqrt{-g}}\frac{\partial}{\partial
x^{\nu}}\left(\sqrt{-g}{\mathcal Z}^{\mu\nu}\right)=0,
\end{equation}
which means that $ {\mathcal
Z}^{\mu\nu}=\frac{1}{\sqrt{-g}}C^{\mu\nu} $ with $C^{\mu\nu}$
being a certain constant tensor, or a tensor whose divergence
vanishes. Because of Eq.(\ref{eq521}), $C^{\mu\nu}=0$, {\it i.e.},
${\mathcal Z}^{\mu\nu}=0$. Note that even though the equation
${\mathcal Z}^{\mu\nu}=0$ can be derived straightforwardly from
Eq.(\ref{eq521}), here we would not think of it as the result of
Eq.(\ref{eq521}). On the contrary, we prefer to consider it as the
result of the covariant divergence of the field equation
Eq.(\ref{eq521}). This, therefore, implies that the equation
${\mathcal Z}^{\mu\nu}=0$ is inevitably in connection with the
kinematic equation of a system of particles consisting of
gravitomagnetic charges. The following calculation will confirm
this interpretation: $S^{\mu\nu}U_{\nu}$ in ${\mathcal
Z}^{\mu\nu}U_{\nu}$ is expressed by
\begin{eqnarray}
S^{\mu\nu}U_{\nu}&=&\kappa_{1}U_{\nu}\left({\rm D}^{\mu}U^{\nu}-{\rm D}^{\nu}U^{\mu}\right)+\frac{1}{2}\kappa_{2}\epsilon^{\mu\nu\alpha\beta}U_{\nu}\left(\partial_{\alpha}U_{\beta}-\partial_{\beta}U_{\alpha}\right)                                                   \nonumber \\
&=&-\kappa_{1}\frac{{\rm {\rm D}} U^{\mu}}{{\rm
d}s}+\frac{1}{2}\kappa_{2}\epsilon^{\mu\nu\alpha\beta}U_{\nu}\left[\partial_{\alpha}\left(g_{\beta\lambda}U^{\lambda}\right)-\partial_{\beta}\left(g_{\alpha\lambda}U^{\lambda}\right)\right]
\nonumber \\
&=&-\kappa_{1}\frac{{\rm  D} U^{\mu}}{{\rm
d}s}+\frac{1}{2}\kappa_{2}\epsilon^{\mu\nu\alpha\beta}U_{\nu}\left(\partial_{\alpha}g_{\beta\lambda}-\partial_{\beta}g_{\alpha\lambda}\right)U^{\lambda}+\frac{1}{2}\kappa_{2}\epsilon^{\mu\nu\alpha\beta}U_{\nu}\left(g_{\beta\lambda}\partial_{\alpha}-g_{\alpha\lambda}\partial_{\beta}\right)U^{\lambda}.
\label{eq524}
\end{eqnarray}
It is noted that the second term
$\frac{1}{2}\kappa_{2}\epsilon^{\mu\nu\alpha\beta}U_{\nu}\left(\partial_{\alpha}g_{\beta\lambda}-\partial_{\beta}g_{\alpha\lambda}\right)U^{\lambda}$
on the right-handed side of (\ref{eq524}) contains an expression
$\sim
\frac{1}{2}\kappa_{2}\epsilon^{\mu\nu\alpha\beta}\left(\partial_{\alpha}g_{\beta
0}-\partial_{\beta}g_{\alpha 0}\right)U_{\nu}$, which is just a
gravitational (gravitomagnetic) Lorentz force, the electromagnetic
counterpart of which is
$g\epsilon^{\mu\nu\alpha\beta}\left(\partial_{\alpha}A_{\beta}-\partial_{\beta}A_{\alpha}\right)U_{\nu}$
with $g$ representing the magnetic charge. The fact that $\sim
\frac{1}{2}\kappa_{2}\epsilon^{\mu\nu\alpha\beta}\left(\partial_{\alpha}g_{\beta
0}-\partial_{\beta}g_{\alpha 0}\right)U_{\nu}$ resembles the
expression for the Lorentz force in electrodynamics means that the
equation ${\mathcal Z}^{\mu\nu}=0$ contains the dynamical equation
of motion of a particle system composed of gravitomagnetic
charges.

The antisymmetric field equation (\ref{eq521}) of gravitomagnetic
matter governs the gravitational field near the gravitomagnetic
monopoles. It does not introduce new extra gravitational
potentials. The role of this equation presented here is to
determine the nonanalytic parts of the metric which is caused by
the existence of the gravitomagnetic monopoles. If the
gravitomagnetic charge does not exist in the universe, the vacuum
case of Eq.(\ref{eq521}) is just the Bianchi identity of
Einstein's field equation.

\subsection{About NUT metric}
The NUT metric is the one produced by the stationary
gravitomagnetic monopole at the origin of the coordinate system.
Here we will discuss the relation between the NUT parameter
$\iota$ and the gravitomagnetic charge. The contravariant-form
field equation (\ref{eq521}) of dual matter can be rewritten as
\begin{equation}
\tilde{{\mathcal
R}}^{\mu\nu}=\frac{1}{\sqrt{-g}}\Upsilon^{\mu\nu}, \quad
S^{\mu\nu}=\frac{1}{\sqrt{-g}}\left(\Upsilon^{\mu\nu}-\Upsilon^{\nu\mu}\right).
\label{eqform}
\end{equation}
For a single gravitomagnetic monopole, the strength of which is
$\iota'$, one may assume that $\Upsilon^{0}_{\ \ 0}=\iota' \delta
(x)$. The four-dimensional volume integral of the term on the
right-handed side of the field equation $\tilde{{\mathcal
R}}^{0}_{\ \ 0 }=\frac{1}{\sqrt{-g}}\Upsilon^{0}_{\ \ 0}$ is
\begin{equation}
\int_{\Omega}\sqrt{-g}\left(\frac{1}{\sqrt{-g}}\Upsilon^{0}_{\ \
0}\right){\rm d}\Omega=\iota'.
\end{equation}
In what follows, we will consider the equation $\tilde{{\mathcal
R}}^{0}_{\ \ 0 }=\frac{1}{\sqrt{-g}}\iota' \delta (x)$, where
$\tilde{{\mathcal R}}^{0}_{\ \ 0
}=\frac{1}{4}\epsilon^{0\lambda\sigma\tau}\partial_{\tau}\left(\partial_{\sigma}g_{0\lambda}-\partial_{\lambda}g_{\sigma
0}\right)$ and
$\epsilon^{0\lambda\sigma\tau}=\frac{1}{\sqrt{-g}}\varepsilon^{0\lambda\sigma\tau}$.
Here $\varepsilon^{0\lambda\sigma\tau}$ is the Levi-Civita tensor
in the flat Minkowski spacetime. It is readily verified that this
equation can be rewritten as

\begin{equation}
\nabla\cdot{\bf B}_{\rm g}=4\iota''\delta({\bf x}), \label{nuteq1}
\end{equation}
where $\iota''=\iota'\delta(x^{0})$, $\delta (x)=\delta
(x^{0})\delta ({\bf x})$. Here the gravitomagnetic field strength
is defined as ${\bf B}_{\rm g}=\nabla\times {\bf g}$ with the
gravitomagnetic vector potentials being ${\bf g}=\left(g_{01},
g_{02}, g_{03}\right)$.

It is well known that the gravitomagnetic vector potentials of NUT
metric (without mass) can be expressed as
\begin{equation}
{\bf g}=\left(0, 0, -2f^{2}(r)\frac{\iota(1-\cos \theta)}{r\sin
\theta}\right)
\end{equation}
in the spherical polar coordinate system, namely,
$g_{0\varphi}=-2f^{2}(r)\frac{\iota(1-\cos \theta)}{r\sin
\theta}$, where the function $f^{2}(r)$ is
$f^{2}(r)=1-2\left(\frac{mr+\iota^{2}}{r^{2}+\iota^{2}}\right)$.
Thus it is easy to obtain
\begin{equation}
\nabla\times {\bf g}=\frac{-2\iota f^{2}(r)}{r^{2}}{\bf
e}_{\rho}+\frac{2}{r}\frac{\partial f^{2}(r)}{\partial
r}\frac{\iota(1-\cos \theta)}{r\sin \theta}{\bf e}_{\theta}.
\end{equation}
It should be noted that the divergence $\nabla\cdot
\left(\nabla\times {\bf g}\right)$ no longer vanishes at some
spatial points (singularities) since here the gravitomagnetic
vector potential is nonanalytic, which results from the presence
of the gravitomagnetic monopoles. Hence the three-dimensional
volume integral of $\nabla\cdot\left(\nabla\times {\bf g}\right)$
is

\begin{equation}
\int_{V}\nabla\cdot\left(\nabla\times {\bf g}\right){\rm d}V=
\mathop{{\int\!\!\!\!\!\int}\mkern-21mu \bigcirc}\limits_{\bf
S}\left(\nabla\times {\bf g}\right)\cdot{\rm d}{\bf S}=-8\pi
\iota. \label{nuteq2}
\end{equation}
It follows from Eq.(\ref{nuteq1}) and (\ref{nuteq2}) that the
relation between $\iota''$ and the NUT parameter $\iota$ is given
\begin{equation}
\iota''=-2\pi \iota.
\end{equation}
It should be emphasized that the Schwarzschild solution also
satisfies the field equation (\ref{eq521}) without source term,
{\it i.e.}, $\tilde{G}^{\mu\nu}=0$. Someone may thus argue that
the Schwarzschild solution is just the empty-space generalization
of the solutions of Eq.(\ref{eq521}). This situation is similar to
the case of the NUT solution $g_{\mu\nu}(m=0, \iota\neq 0)$, which
was also thought of as the empty-space generalization of the
solutions of Einstein's equation\cite{Newman}.

How can we look upon the two solutions $g_{\mu\nu}(m=0, \iota\neq
0)$ and $g_{\mu\nu}(m\neq 0, \iota=0)$ in the NUT spacetime? It is
believed that the vacuum equation ($\tilde{G}^{\mu\nu}=0$) of dual
matter may be viewed as the Bianchi identity of Einstein's field
equation, and Einstein's vacuum equation is in turn the Bianchi
identity of the field equation (\ref{eq521}) of gravitomagnetic
matter. For this reason, even though the solution
$g_{\mu\nu}(m\neq 0, \iota=0)$ is truly the one to Einstein's
field equation, it can also agree with the equation
$\tilde{G}^{\mu\nu}=0$ (Bianchi identity). So, it may be argued
that the solution $g_{\mu\nu}(m\neq 0, \iota=0)$ is the
empty-space generalization of the solution of vacuum equation
$\tilde{G}^{\mu\nu}=0$. Likewise, the reason for Newman {\it et
al.} to think of $g_{\mu\nu}(m=0, \iota\neq 0)$ as the solution of
Einstein's vacuum equation\cite{Newman} lies in that Einstein's
vacuum equation is just the Bianchi identity of Eq.(\ref{eq521})
of gravitomagnetic matter. In fact, $g_{\mu\nu}(m=0, \iota\neq 0)$
is only the one of the solutions of Eq.(\ref{eq521}), and
furthermore, all the solutions satisfying Eq.(\ref{eq521}) will
also agree with Einstein's vacuum equation (Bianchi identity). In
this sense, the gravitomagnetic charge is truly required of its
own gravitational field equation for us to treat its dynamical
problems.

In addition, we consider briefly the stationary cylindrically
symmetric exact solution of field equation (\ref{nuteq1}). Suppose
that the form of linear element describing the stationary
cylindrically symmetric gravitomagnetic field is given by
\begin{equation}
{\rm d}s^{2}=({\rm d}{x^{0}})^{2}-{\rm d}x^{2}-{\rm d}y^{2}-{\rm d}z^{2}+2g_{0x}(y)%
{\rm d}x^{0}{\rm d}x+2g_{0y}(x){\rm d}x^{0}{\rm d}y,
\label{cylineq20}
\end{equation}
where we assume that the gravitomagnetic potentials $g_{0x}$ and
$g_{0y}$ are the functions with respect to $y$ and $x$,
respectively. For the cylindrically symmetric uniform
gravitomagnetic field produced by the gravitomagnetic charges
(with a nonvanishing surface density of $\sigma _{M}$) present
only in the $x$-$y$ plane of $z=0$, the gravitomagnetic field,
${\bf B}_{g}$, which is defined as $\nabla \times {\bf g}$, may be
written
\begin{equation}
({\bf B}_{g})_{z}=\frac{\sigma _{M}}{2}.    \label{cylineq27}
\end{equation}
It follows from Eq.(\ref{cylineq27}) that the direction of ${\bf
B}_{g}$ is parallel to the $z$-axis. The metric components of the
uniform gravitomagnetic field, $g_{0x}$, $g_{0y}$, are therefore
readily obtained as follows
\begin{equation}
g_{0x}=\frac{B_{g}}{2}y,\quad g_{0y}=-\frac{B_{g}}{2}x,
\label{cylineq28}
\end{equation}
with $B_{g}=\frac{\sigma _{M}}{2}$.

In order to obtain the contravariant metric $g^{\mu \nu }$, we
calculate the inverse matrix of the metric $\left( g_{\mu \nu
}\right)$, and the result is given as follows
\begin{equation}
\left( g^{\mu \nu }\right)
=\frac{1}{1+g_{0x}^{2}+g_{0y}^{2}}\left(
\begin{array}{cccc}
{1} & {g_{0x}} & {g_{0y}} & {0} \\
{g_{0x}} & {-(1+g_{0y}^{2})} & {g_{0x}g_{0y}} & {0} \\
{g_{0y}} & {g_{0x}g_{0y}} & {-(1+g_{0x}^{2})} & {0} \\
{0} & {0} & {0} & {-(1+g_{0x}^{2}+g_{0y}^{2})}
\end{array}
\right).  \label{cylineq29}
\end{equation}

\subsection{Dual spin connection}
Here we will consider the covariant derivative of Dirac field
composed of gravitomagnetic charge in the local frame. Assume that
an observer moves with proper three-acceleration ${\bf a}$ and
proper three-rotation $\vec{\omega}$ inside an inertial frame of
reference. Note that here the quantities and symbols just refer to
those in the reference\cite{Ni}. It is believed that the covariant
derivative of gravitomagnetic-charge Dirac field is given as
follows
\begin{equation}
D_{\hat{\alpha}}=\partial_{\hat{\alpha}}-\frac{i}{16}\sigma^{\hat{\beta}\hat{\gamma}}\varepsilon_{\hat{\beta}\hat{\gamma}}^{\
\ \
\hat{\lambda}\hat{\tau}}\Gamma_{\hat{\lambda}\hat{\tau}\hat{\alpha}}.
\end{equation}
Now we discuss the dual spin connection
$-\frac{i}{16}\sigma^{\hat{\beta}\hat{\gamma}}\varepsilon_{\hat{\beta}\hat{\gamma}}^{\
\ \
\hat{\lambda}\hat{\tau}}\Gamma_{\hat{\lambda}\hat{\tau}\hat{\alpha}}$.
By using the following relations\cite{Ni}
\begin{eqnarray}
\Gamma_{\hat{i}\hat{j}\hat{0}}&=&-
\frac{\frac{\varepsilon_{ijk}\omega^{k}}{c}}{1+\frac{{\bf
a}\cdot{\bf x}}{c^{2}}},  \quad
\Gamma_{\hat{0}\hat{l}\hat{0}}=-\Gamma_{\hat{l}\hat{0}\hat{0}}=-\frac{\frac{a^{i}}{c^{2}}}{1+\frac{{\bf
a}\cdot{\bf
x}}{c^{2}}},  \quad        \Gamma_{\hat{\mu}\hat{\nu}\hat{0}}=\Gamma_{\hat{0}\hat{0}\hat{0}}=0,          \nonumber \\
\frac{1}{2}\sigma^{ij}&=&-\frac{{\bf S}}{\hbar},    \quad
i\sigma^{0l}=\vec{\alpha},
\end{eqnarray}
one can verify that
\begin{eqnarray}
-\frac{i}{16}\left(\sigma^{bc}\varepsilon_{bc}^{\ \ \
ij}\Gamma_{ij0}+2\sigma^{bc}\varepsilon_{bc}^{\ \ \
0l}\Gamma_{0l0}\right) &=& \frac{1}{1+\frac{{\bf a}\cdot{\bf
x}}{c^{2}}}\left(\frac{i}{16}\sigma^{bc}\varepsilon_{bc}^{\ \ \
ij}\varepsilon_{ijk}\frac{\omega^{k}}{c}+\frac{i}{8}\sigma^{bc}\varepsilon_{bc}^{\
\ \ 0l}\frac{a^{l}}{c^{2}}\right)                  \nonumber \\
&=& \frac{1}{1+\frac{{\bf a}\cdot{\bf
x}}{c^{2}}}\left(\frac{1}{4c}\vec{\omega}\cdot\vec{\alpha}-\frac{i}{2\hbar
c^{2}}{\bf a}\cdot{\bf S}\right),      \label{eqvelocity}
\end{eqnarray}
where the Latin indices run over $1, 2, 3$. Here $\vec{\alpha}$
and ${\bf S}$ denote the velocity operator and spin operator of
Dirac field, respectively.

It is well known that a Dirac particle in a noninertial frame of
reference will undergo the spin-rotation coupling
($\vec{\omega}\cdot{\bf S}$)\cite{Mashhoon1,Mashhoon2} and
velocity-acceleration coupling (${\bf
a}\cdot\vec{\alpha}$)\cite{Ni}. However, it follows from
(\ref{eqvelocity}) that for a gravitomagnetic-charge Dirac field,
it experiences velocity-rotation coupling
($\vec{\omega}\cdot\vec{\alpha}$) and spin-acceleration coupling
(${\bf a}\cdot{\bf S}$), which are just the dual interactions of
the above two couplings.

\section{Dual metric and dual current densities}
\subsection{Dual vierbein field and dual metric}
It can be shown that in the weak gravitational field, Dirac
matrices $\gamma^{\mu}$'s is rewritten as
$\gamma^{0}=\left(1+\frac{\phi}{c^{2}}\right)\beta$,
$\gamma^{i}=g^{0i}\beta+\gamma^{i}_{\rm M}$ ($i=1, 2,
3$)\cite{Shen}. Here $\beta$ and $\gamma^{i}_{\rm M}$ are the
Dirac matrices in the flat Minkowski spacetime. $\phi$ and
$g^{0i}$ can be viewed as the gravitational scalar potential and
gravitomagnetic vector potentials. By using $\gamma^{\mu}=e^{\mu
a}\gamma_{a}$ ($a$ denotes the indices of coordinate in Minkowski
spacetime), one can obtain the relationship between the vierbein
field $e^{\mu a}$ and the gravitational potential and
gravitomagnetic vector potentials: $e^{00}\simeq
1+\frac{\phi}{c^{2}}$ and $e^{\mu 0}\simeq g^{\mu 0}$ ($\mu=1, 2,
3$). Here the dual vierbein field $\tilde{e}_{\mu a}$ is so
defined that $\tilde{e}_{\mu a}$ and $\tilde{e}_{\nu a}$ satisfy
the following relation
\begin{equation}
\partial_{\mu}\tilde{e}_{\nu a}-\partial_{\nu}\tilde{e}_{\mu
a}=\frac{1}{2}\epsilon_{\mu\nu}^{\ \ \
\alpha\beta}\left(\partial_{\alpha}e_{\beta
a}-\partial_{\beta}e_{\alpha a}\right). \label{eqdefinition}
\end{equation}
Moreover, the dual Dirac matrices can be defined as
$\tilde{\gamma}_{\mu}=\tilde{e}_{\mu a}\gamma^{a}$. Multiplying
the two sides of the above equation by $\gamma^{a}$, one can
arrive at
\begin{equation}
\partial_{\mu}\tilde{\gamma}_{\nu}-\partial_{\nu}\tilde{\gamma}_{\mu}=\frac{1}{2}\epsilon_{\mu\nu}^{\ \
\
\alpha\beta}\left(\partial_{\alpha}\gamma_{\beta}-\partial_{\beta}\gamma_{\alpha}\right),
\end{equation}
which is the relation between the dual Dirac matrices
$\tilde{\gamma}_{\mu}$ and the regular Dirac matrices
$\gamma_{\alpha}$.

The connection between the metric $g_{\mu\nu}$ and the flat metric
$\eta_{ab}$ is $g_{\mu\nu}=e_{\mu}^{\ a}e_{\nu}^{\ b}\eta_{ab}$.
Correspondingly, the relation between the dual metric
$\tilde{g}_{\mu\nu}$ and the flat metric $\eta_{ab}$ may be
$\tilde{g}_{\mu\nu}=\tilde{e}_{\mu}^{\ a}\tilde{e}_{\nu}^{\
b}\eta_{ab}$. It should be noted that even though we introduce the
dual metric $\tilde{g}_{\mu\nu}$, in general, indices of tensors
are raised and lowered still with the metric tensor ${g}_{\mu\nu}$
rather than with the dual metric $\tilde{g}_{\mu\nu}$.

In this subsection, we will discuss the dual covariant
derivatives, $\tilde{\nabla}_{\mu}\psi$, of Fermionic field
$\psi$\cite{Gao}. The dual spin connection of
$\tilde{\nabla}_{\mu}\psi=\left(\partial_{\mu}+i\tilde{B}_{\mu}\right)\psi$
is defined by
\begin{equation}
\tilde{B}_{\mu}=\frac{1}{16}e_{\beta c}e^{\beta}_{\ \
a;\mu}\varepsilon^{ca}_{\ \ \ bd}\sigma^{bd},
\end{equation}
where $\sigma^{bd}=\frac{i}{2}\left[\gamma^{b},
\gamma^{d}\right]$. It is essentially significant to construct the
Lagrangian density of Fermionic-field gravitomagnetic charge,
which may be given
\begin{equation}
{\mathcal L}_{\rm F}=\sqrt{-g}\left\{\frac{i}{2}\tilde{e}^{\mu
a}\left[\bar{\psi}\gamma_{a}\tilde{\nabla}_{\mu}\psi-\left(\tilde{\nabla}_{\mu}\bar{\psi}\right)\gamma_{a}\psi\right]-\tilde{m}\bar{\psi}\psi\right\},
\end{equation}
where $\tilde{m}$ denotes the dual mass of gravitomagnetic
monopole, and
$\tilde{\nabla}_{\mu}\bar{\psi}=\left(\partial_{\mu}-i\tilde{B}_{\mu}\right)\bar{\psi}$.
The corresponding classical field equations of $\psi$ and
$\bar{\psi}$ are as follows
\begin{eqnarray}
\frac{i}{2}\tilde{e}^{\mu a}\gamma_{a}\tilde{\nabla}_{\mu}\psi+\frac{i}{2}\frac{1}{\sqrt{-g}}\partial_{\mu}\left(\sqrt{-g}\tilde{e}^{\mu a}\gamma_{a}\psi\right)-\tilde{m}\psi=0,                 \nonumber \\
-\frac{i}{2}\frac{1}{\sqrt{-g}}\partial_{\mu}\left(\sqrt{-g}\bar{\psi}\tilde{e}^{\mu
a}\gamma_{a}\right)-\frac{i}{2}\left(\tilde{\nabla}_{\mu}\bar{\psi}\right)\tilde{e}^{\mu
a}\gamma_{a}-\tilde{m}\bar{\psi}=0.       \label{fermion}
\end{eqnarray}
It follows from Eqs.(\ref{fermion}) that the conservation law of
the dual current density is
\begin{equation}
\frac{1}{\sqrt{-g}}\partial_{\mu}\left(\sqrt{-g}i\bar{\psi}\tilde{e}^{\mu
a}\gamma_{a}\psi\right)=0,
\end{equation}
where the dual current density of dual mass is
\begin{equation}
\tilde{j}^{\mu}=i\bar{\psi}\tilde{e}^{\mu
a}\gamma_{a}\psi=\tilde{e}^{\mu a}j_{a},   \label{dualcurrent}
\end{equation}
which will be considered further in the following.

In addition, the Lagrangian density of the Bosonic-field
gravitomagnetic matter may be of the form
\begin{equation}
{\mathcal L}_{\rm B}=\sqrt{-g}\left[\tilde{e}^{\mu
a}\left(\partial_{\mu}\varphi^{\ast}\partial_{a}\varphi+\partial_{a}\varphi^{\ast}\partial_{\mu}\varphi\right)-\tilde{m}^{2}\varphi^{\ast}\varphi\right].
\end{equation}
The corresponding classical field equations are
\begin{eqnarray}
\frac{1}{\sqrt{-g}}\partial_{\mu}\left(\sqrt{-g}\tilde{e}^{\mu a}\partial_{a}\varphi\right)+\tilde{m}^{2}\varphi=0,                  \nonumber \\
\frac{1}{\sqrt{-g}}\partial_{\mu}\left(\sqrt{-g}\tilde{e}^{\mu
a}\partial_{a}\varphi^{\ast}\right)+\tilde{m}^{2}\varphi^{\ast}=0.
\end{eqnarray}
It follows that the conservation law of the dual current density
takes the form
\begin{equation}
\frac{1}{\sqrt{-g}}\partial_{\mu}\left[\sqrt{-g}\tilde{e}^{\mu
a}\left(\varphi^{\ast}\partial_{a}\varphi-\varphi\partial_{a}\varphi^{\ast}\right)\right]=0.
\end{equation}
Thus the dual current density of Bosonic-field gravitomagnetic
matter can be written as
\begin{equation}
\tilde{j}^{\mu}=\frac{1}{2i}\tilde{e}^{\mu
a}\left(\varphi^{\ast}\partial_{a}\varphi-\varphi\partial_{a}\varphi^{\ast}\right)=\tilde{e}^{\mu
a}j_{a}.
\end{equation}
Note that here the index $a$ refers to the one in the flat
Minkowski spacetime.

It is shown that in the above classical field equations of
Fermionic and Bosonic field, gravitomagnetic monopole possesses
the dual mass rather than the mass, namely, the concept of the
mass is of no significance for the gravitomagnetic matter. Dual
mass has its own gravitational nature and features different much
from that of the mass. In this sense, matter may be classified
into two categories, {\it i.e.}, the regular matter and dual
matter. The former has {\it mass} while the latter has {\it dual
mass}.

\subsection{Dual current densities}
In accordance with the expression (\ref{dualcurrent}), the dual
velocity (or current density) of dual mass can be defined as
$\tilde{U}_{\mu}=\tilde{e}_{\mu}^{\ a}U_{a}$. To compare
$\partial_{\mu}\tilde{U}_{\nu}-\partial_{\nu}\tilde{U}_{\mu}$ with
$\partial_{\mu}{U}_{\nu}-\partial_{\nu}{U}_{\mu}$ is physically
interesting, which is valuable for us to define a new kind of the
dual vierbein field. The calculation of
$\partial_{\mu}\tilde{U}_{\nu}-\partial_{\nu}\tilde{U}_{\mu}$ and
$\partial_{\mu}{U}_{\nu}-\partial_{\nu}{U}_{\mu}$ yields
\begin{eqnarray}
&&
\partial_{\mu}\tilde{U}_{\nu}-\partial_{\nu}\tilde{U}_{\mu}=\left[\partial_{\mu}\tilde{e}_{\nu
a}-\partial_{\nu}\tilde{e}_{\mu a}-\left(\tilde{e}_{\mu
a}\partial_{\nu}-\tilde{e}_{\nu
a}\partial_{\mu}\right)\right]U^{a},     \nonumber   \\
&&
\partial_{\mu}{U}_{\nu}-\partial_{\nu}{U}_{\mu}=\left[\partial_{\mu}{e}_{\nu
a}-\partial_{\nu}{e}_{\mu a}-\left({e}_{\mu
a}\partial_{\nu}-{e}_{\nu a}\partial_{\mu}\right)\right]U^{a}.
\label{calculation}
\end{eqnarray}
Here we will adopt an alternative definition of the dual vierbein
field $\tilde{e}_{\mu a}$, {\it i.e.},
\begin{equation}
\partial_{\mu}\tilde{e}_{\nu a}-\partial_{\nu}\tilde{e}_{\mu
a}-\left(\tilde{e}_{\mu a}\partial_{\nu}-\tilde{e}_{\nu
a}\partial_{\mu}\right)=\frac{1}{2}\epsilon_{\mu\nu}^{\ \ \
\alpha\beta}\left[\partial_{\alpha}{e}_{\beta
a}-\partial_{\beta}{e}_{\alpha a}-\left({e}_{\alpha
a}\partial_{\beta}-{e}_{\beta a}\partial_{\alpha}\right)\right],
\label{eqdefine1}
\end{equation}
which is different from (\ref{eqdefinition}). Consequently, this
definition will lead to the following duality relation
\begin{equation}
\partial_{\mu}{e}_{\nu a}-\partial_{\nu}{e}_{\mu
a}-\left({e}_{\mu a}\partial_{\nu}-{e}_{\nu
a}\partial_{\mu}\right)=-\frac{1}{2}\epsilon_{\mu\nu}^{\ \ \
\alpha\beta}\left[\partial_{\alpha}\tilde{e}_{\beta
a}-\partial_{\beta}\tilde{e}_{\alpha a}-\left(\tilde{e}_{\alpha
a}\partial_{\beta}-\tilde{e}_{\beta
a}\partial_{\alpha}\right)\right].             \label{eqdefine2}
\end{equation}
Thus, it follows from Eq.(\ref{eqdefine1}) and (\ref{eqdefine2})
that Eq.(\ref{calculation}) can be rewritten as
$\partial_{\mu}\tilde{U}_{\nu}-\partial_{\nu}\tilde{U}_{\mu}=\frac{1}{2}\epsilon_{\mu\nu}^{\
\ \
\alpha\beta}\left(\partial_{\alpha}{U}_{\beta}-\partial_{\beta}{U}_{\alpha}\right)$
and
$\partial_{\mu}{U}_{\nu}-\partial_{\nu}{U}_{\mu}=-\frac{1}{2}\epsilon_{\mu\nu}^{\
\ \
\alpha\beta}\left(\partial_{\alpha}\tilde{U}_{\beta}-\partial_{\beta}\tilde{U}_{\alpha}\right)$.
So, the source tensors ${\mathcal K}_{\mu\nu}$ and ${\mathcal
H}_{\mu\nu}$ defined in Sec. V can also be expressed in terms of
the dual vierbein field (or dual current density), {\it i.e.},
there is a duality relation between ${\mathcal K}_{\mu\nu}$,
${\mathcal H}_{\mu\nu}$ and $\tilde{{\mathcal H}}_{\mu\nu}$,
$\tilde{{\mathcal K}}_{\mu\nu}$:
\begin{eqnarray}
& & {\mathcal
K}_{\mu\nu}\equiv\partial_{\mu}{U}_{\nu}-\partial_{\nu}{U}_{\mu}=-\frac{1}{2}\epsilon_{\mu\nu}^{\
\ \
\alpha\beta}\left(\partial_{\alpha}\tilde{U}_{\beta}-\partial_{\beta}\tilde{U}_{\alpha}\right)\equiv
-\tilde{{\mathcal H}}_{\mu\nu},               \nonumber \\
& &       {\mathcal H}_{\mu\nu}\equiv
\frac{1}{2}\epsilon_{\mu\nu}^{\ \ \
\alpha\beta}\left(\partial_{\alpha}{U}_{\beta}-\partial_{\beta}{U}_{\alpha}\right)=\partial_{\mu}\tilde{U}_{\nu}-\partial_{\nu}\tilde{U}_{\mu}\equiv
\tilde{{\mathcal K}}_{\mu\nu}.
\end{eqnarray}

The above discussion shows that if the vierbein field
$\tilde{e}_{\mu a}$ is defined via Eq.(\ref{eqdefine1}), then
there is no difference between the source tensor $S_{\mu\nu}$ and
the dual source tensor $\tilde{S}_{\mu\nu}$. Here the dual source
tensor $\tilde{S}_{\mu\nu}$ is the linear combination of
$\tilde{{\mathcal K}}_{\mu\nu}$ and $\tilde{{\mathcal
H}}_{\mu\nu}$. In a word, both the current density and the dual
current density can be used to define the source tensor of
gravitomagnetic matter in the gravitational field equation.

\section{The five-dimensional case}
In order to find the potential connection between the
gravitomagnetic charge and the magnetic charge, we will
investigate the five-dimensional dynamics of gravitomagnetic
matter. In the five-dimensional case, the dual Ricci tensor is
\begin{equation}
\tilde{\hat{{\mathcal R}}}_{AB}=\frac{1}{4}\epsilon_{A}^{\ \
CDE}\frac{\partial}{\partial x^{E}}\left(\frac{\partial
\hat{g}_{BC}}{\partial x^{D}}-\frac{\partial
\hat{g}_{DB}}{\partial x^{C}}\right),
\end{equation}
where $A,B,C,D,E$ run over $0-4$, instead of $0-3$. In what
follows Greek indices run over $0-3$. By using the cylinder
condition ({\it i.e.}, the derivative of the metric with respect
to $x^{4}$ is vanishing), we can obtain
\begin{eqnarray}
& & \tilde{\hat{{\mathcal
R}}}_{\mu\nu}=\frac{1}{4}\epsilon_{\mu}^{\ \
\lambda\sigma\tau}\partial_{\tau}\left(\partial_{\sigma}g_{\nu\lambda}-\partial_{\lambda}g_{\sigma\nu}\right)+\frac{1}{2}\epsilon_{\mu}^{\
\ 4\sigma\tau}\partial_{\tau}\partial_{\sigma}\hat{g}_{\nu 4},
\nonumber   \\
& & \tilde{\hat{{\mathcal R}}}_{\mu
4}=\frac{1}{4}\epsilon_{\mu}^{\ \
\lambda\sigma\tau}\partial_{\tau}\left(\partial_{\sigma}
\hat{g}_{4\lambda}-\partial_{\lambda}\hat{g}_{\sigma
4}\right)+\frac{1}{2}\epsilon_{\mu}^{\ \
4\sigma\tau}\partial_{\tau}\partial_{\sigma}\hat{g}_{44}.
\label{eq72}
\end{eqnarray}
By using the Kaluza mechanism, the five-dimensional metric
$\hat{g}_{AB}$ is adopted as follows\cite{Overduin}
\begin{equation}
(\hat{g}_{AB})=\left(\begin{array}{cccc}
{g_{\alpha\beta}+\kappa^{2}\phi^{2}A_{\alpha}A_{\beta}}  & {\kappa}\phi^{2}A_{\alpha}  \\
{\kappa\phi^{2}A_{\beta}} &   {\phi^{2}}
 \end{array}
 \right),
 \end{equation}
where $\phi$ and $A_{\alpha}$ denote a certain (constant) scalar
field and the electromagnetic field, respectively. Thus
Eqs.(\ref{eq72}) can be rewritten as
\begin{eqnarray}
\tilde{\hat{{\mathcal R}}}_{\mu\nu}=\tilde{{{\mathcal
R}}}_{\mu\nu}+\frac{\kappa\phi^{2}}{2}\epsilon_{\mu}^{\ \
4\sigma\tau}\partial_{\tau}\partial_{\sigma}A_{\nu},     \quad
\tilde{\hat{{\mathcal R}}}_{\mu
4}=\frac{\kappa\phi^{2}}{4}\epsilon_{\mu}^{\ \
\lambda\sigma\tau}\partial_{\tau}{\mathcal F}_{\sigma\lambda}.
\label{second}
\end{eqnarray}
So according to the field equation (\ref{eqform}), {\it i.e.},
$\tilde{\hat{{\mathcal R}}}_{\mu
4}=\frac{1}{\sqrt{-g}}\hat{\Upsilon}_{\mu 4}$, the second equation
in Eqs.(\ref{second}) can be rewritten as
\begin{equation}
\frac{\kappa\phi^{2}}{4}\epsilon_{\mu}^{\ \
\lambda\sigma\tau}\partial_{\tau}{\mathcal
F}_{\sigma\lambda}=\frac{1}{\sqrt{-g}}\hat{\Upsilon}_{\mu 4},
\end{equation}
which is actually the electromagnetic field equation of magnetic
charge. This, therefore, means that the $(\mu 4)$ component of the
five-dimensional gravitational field equation of gravitomagnetic
matter can be reduced to the electromagnetic field equation of
magnetic monopole via the Kaluza mechanism. Thus the
five-dimensional dynamics of gravitomagnetic monopole provides us
with a potentially unified way to treat the magnetic monopole and
gravitomagnetic monopole.

\section{Dual theory}
In electrodynamics, it is shown that the dynamics of magnetic
charge is just the dual theory to that of electric charge (for a
concise formalism, see Appendix 6 to this paper). This phenomenon
arises also in gravity theory. Now we are therefore led to
consider in detail the duality relationship between the dynamical
theories of gravitomagnetic charge (dual mass) and gravitoelectric
charge (mass). First we define a second-rank antisymmetric
gravitational potential tensor $h^{\mu\nu}$ as an alternative to
the description of the gravitational field. Such antisymmetric
gravitational potential tensor $h^{\mu\nu}$ may be so defined that
the following expression
\begin{equation}
\epsilon^{\mu\tau\omega\nu}=\sigma(x)\left(h^{\mu\omega}h^{\tau\nu}+h^{\tau\mu}h^{\omega\nu}+h^{\nu\mu}h^{\tau\omega}\right)
\label{eq61}
\end{equation}
is satisfied, where $\sigma(x)$ is a scalar function to be
determined.

Define a functional of $h^{\lambda\sigma}$, say, ${\mathcal
N}_{\tau\omega\nu\mu}\left(h^{\lambda\sigma}\right)$, which is
assumed to contain the derivatives of $h^{\lambda\sigma}$ up to
second order ({\it i.e.}, it contains
$\partial_{\theta}h^{\lambda\sigma}$, $\partial_{\theta
\eta}h^{\lambda\sigma}$)\footnote{It is assumed that the field
equations associated with the gravitational field is often of
second differential order.}. Based on this, we assume that the
gravitational Lagrangian density is chosen to be
\begin{equation}
{\mathcal
L}=-\frac{1}{2}\sqrt{-g}\epsilon^{\mu\tau\omega\nu}{\mathcal
N}_{\tau\omega\nu\mu}\left(h^{\lambda\sigma}\right). \label{eq62}
\end{equation}
By using the Euler-Lagrange equation, if the following equation is
satisfied
\begin{equation}
\frac{\partial {\mathcal L}}{\partial
h^{\lambda\sigma}}-\left[\frac{\partial {\mathcal L}}{\partial
h^{\lambda\sigma}_{\ \
,\theta}}\right]_{,\theta}+\left[\frac{\partial {\mathcal
L}}{\partial h^{\lambda\sigma}_{\ \
,\theta\eta}}\right]_{,\theta\eta}=\sqrt{-g}g^{\omega\nu}\left({\mathcal
N}_{\lambda\omega\nu\sigma}-{\mathcal
N}_{\sigma\omega\nu\lambda}\right),
\label{eq63}
\end{equation}
then it is believed that the functional ${\mathcal
N}_{\tau\omega\nu\mu}\left(h^{\lambda\sigma}\right)$ is related
closely to the dual Riemann tensor $\tilde{{\mathcal
R}}_{\tau\omega\nu\mu}$, namely, keeping in mind Eq.(\ref{eq62}),
we can set
\begin{equation}
\tilde{{\mathcal R}}_{\tau\omega\nu\mu}={\mathcal
N}_{\tau\omega\nu\mu}\left(h^{\lambda\sigma}\right),
\label{eq64}
\end{equation}
which results from the expression (\ref{eq42}),
$\frac{1}{2}\epsilon^{\mu\tau\omega\nu}\tilde{{\mathcal
R}}_{\tau\omega\nu\mu}=-{\mathcal R}$, in Sec. IV.

The roles of Eq.(\ref{eq61}), (\ref{eq63}) and (\ref{eq64}) is as
follows: Eq.(\ref{eq61}) can be used to determine the scalar
function $\sigma(x)$; Eq.(\ref{eq63}) is applied to the
determination of the form of the functional ${\mathcal
N}_{\tau\omega\nu\mu}\left(h^{\lambda\sigma}\right)$;
Eq.(\ref{eq64}) can be employed to find the relationship between
the antisymmetric gravitational potential tensor
$h^{\lambda\sigma}$ and the symmetric metric tensor
$g^{\lambda\sigma}$.

Because of the identity,
$\frac{1}{2}\epsilon^{\theta\tau\omega\nu}\tilde{{\mathcal
R}}^{\mu}_{\ \
\tau\omega\nu}=-\frac{1}{2}\left(G^{\theta\mu}+G'^{\theta\mu}\right)$,
suggested in Sec. III, the substitution of Eq.(\ref{eq64}) into
this identity leads to
$\frac{1}{2}\left(G^{\theta\mu}+G'^{\theta\mu}\right)=-\frac{1}{2}\epsilon^{\theta\tau\omega\nu}{{\mathcal
N}}^{\mu}_{\ \ \tau\omega\nu}\left(h^{\lambda\sigma}\right)$. Thus
Einstein's field equation is of the following form (rewritten in
terms of the antisymmetric gravitational potential tensor
$h^{\lambda\sigma}$ rather than of the symmetric metric tensor
$g^{\lambda\sigma}$)
\begin{equation}
-\frac{1}{2}\epsilon^{\theta\tau\omega\nu}{\mathcal N}^{\mu}_{\ \
\tau\omega\nu}\left(h^{\lambda\sigma}\right)=-\kappa T^{\theta\mu}
\end{equation}
with $T^{\theta\mu}$ being the energy-momentum tensor of matter.
In this sense, we have found the duality relationship between the
theories of mass and dual mass, and therefore developed a dual
formalism (in terms of the antisymmetric gravitational potential
tensor $h^{\lambda\sigma}$) which can express Einstein's equation
within its framework.

\section{Concluding remarks}
Not all of the nonanalytic solutions of Einstein's vacuum equation
belong to the contribution of gravitomagnetic charge. In an
attempt to investigate the gravitational field caused by
gravitomagnetic charge currents, we meet, however, with
difficulties in selecting the ``true'' metric produced by
gravitomagnetic charge from the solutions of the Bianchi identity
({\it i.e.}, Einstein's vacuum equation). For this reason, one
should have a dynamics of gravitomagnetic charge. To achieve this
aim, this paper gives a concise presentation of the mathematical
formalism of the field equation of gravitomagnetic matter. We
derive the antisymmetric dual Einstein tensor from the variational
principle with the dual curvature scalar, and construct an
antisymmetric source tensor, which appears on the right-handed
side of the gravitational field equation of gravitomagnetic
matter. It may be believed that, as Lynden-Bell and Nouri-Zonoz
stated\cite{Lynden-Bell}, the above-developed gravitational field
equation might introduce a greater degree of physical
understanding of NUT space (and Taub-NUT space).

It is shown in this paper that the dual physical quantities such
as the dual vierbein field (and hence dual metric), dual spin
connection and dual current density (and hence dual source tensor)
constitute the dynamical framework of gravitomagnetic matter. It
is also demonstrated that the topological dual charge needs its
own gravitational field equation. To clarify the relationship
between the field equation of gravitomagnetic matter and
Einstein's field equation is of physical essence. Einstein's field
equation cannot govern independently the spacetime produced by
gravitomagnetic matter. Instead, it can govern independently the
gravitational field of regular matter only. The vacuum equation of
dual matter ({\it i.e.}, without dual sources) is just the Bianchi
identity of Einstein's equation, and similarly, Einstein's vacuum
equation is just the Bianchi identity of the gravitational field
equation of gravitomagnetic matter. The NUT solution is the one to
both Einstein's equation and Eq.(\ref{eq521}).

It follows from the point of view of the classical field equation
that matter may be classified into two categories, {\it i.e.}, the
gravitoelectric and gravitomagnetic matter. It is clear that the
concept of mass is of no significance for the gravitomagnetic
matter. In this sense, we can say that the gravitomagnetic matter
possesses the topological {\it dual mass}\cite{Shen,Shenarxiv}.

Lynden-Bell and Nouri-Zonoz considered briefly the problem as to
how general relativity must be modified if the gravitomagnetic
monopole truly exists in the universe\cite{Lynden-Bell}. They
thought that the space where the gravitomagnetic monopole is
present may possess the unsymmetrical affine
connections\cite{Lynden-Bell}. Namely, their consideration means
that the extra gravitational potentials should be introduced to
allow for the gravitomagnetic charge. Indeed, the topological dual
mass will introduce some additional new things in gravity theory,
{\it e.g.}, it will give rise to the topological singularities in
spacetime, which is described by the nonanalytic metric functions.
But there is no essential reason for cluing us on introducing the
extra gravitational potentials. To deal with the dynamics of
gravitomagnetic charge, the only thing we should do first is to
introduce a new field equation ({\it e.g.}, (\ref{eq5211}) and
(\ref{eq521})), the vacuum case (without the sources) of which is
just the Bianchi identity of Einstein's equation.

Gravitomagnetic charge has some interesting relativistic quantum
gravitational effects, {\it e.g.}, the gravitational anti-Meissner
effect, which  may serve as an interpretation of the smallness of
the observed cosmological constant. In accordance with quantum
field theory, vacuum possesses infinite zero-point energy density
due to the vacuum quantum fluctuations; whereas according to
Einstein's theory of General Relativity, infinite vacuum energy
density yields the divergent curvature of space-time, namely, the
space-time of vacuum is extremely curved. Apparently it is in
contradiction with the practical fact\cite{Weinberg}. In the
context of quantum field theory a cosmological constant
corresponds to the energy density associated with the vacuum and
then the divergent cosmological constant may result from the
infinite energy density of vacuum quantum fluctuations. However, a
diverse set of observations suggests that the universe possesses a
nonzero but very small cosmological
constant\cite{Weinberg,Datta,Krauss,Kakushadze}. How can we give a
natural interpretation for the above paradox? Here, provided that
vacuum matter is perfect fluid, which leads to the formal
similarities between the weak-gravity equation in perfect fluid
and the London's electrodynamics of superconductivity, we suggest
a potential explanation by using the cancelling mechanism via
gravitational anti-Meissner effect: the gravitoelectric field
(Newtonian field of gravity) produced by the gravitoelectric
charge (mass) of the vacuum quantum fluctuations is  exactly
cancelled by the gravitoelectric field due to the induced current
of the gravitomagnetic charge of the vacuum quantum fluctuations;
the gravitomagnetic field produced by the gravitomagnetic charge
(dual mass) of the vacuum quantum fluctuations is exactly
cancelled by the gravitomagnetic field due to the induced current
of the gravitoelectric charge (mass current) of the vacuum quantum
fluctuations. Thus, at least in the framework of weak-field
approximation, the extreme space-time curvature of vacuum caused
by the large amount of the vacuum energy does not arise, and the
gravitational effects of cosmological constant is eliminated by
the contributions of the gravitomagnetic charge (dual mass). If
gravitational anti-Meissner effect is of really physical
significance, then it is necessary to apply this effect to the
early universe where quantum and inflationary cosmologies dominate
the evolution of the universe. Study of the geometric property in
quantum regimes is an interesting and valuable direction.

We think that there might exist the formation/creation mechanism
of gravitomagnetic charge in the gravitational interaction, just
as some prevalent theories provide the theoretical mechanism of
the existence of magnetic monopole in various gauge
interactions\cite{Hooft}. But so far it has to yield any definite
theories of how the gravitomagnetic matter forms (we think that it
may arise from the interactions associated with the Chern-Simons
gauge field). The magnetic monopole in the grand unified theory
plays an important role in the early universe. The same may be
said of gravitomagnetic monopole. If the gravitomagnetic monopoles
truly exist in the universe, it will inevitably give rise to
significant consequences and may also play an essential role in
astrophysics and cosmology. In this paper we investigated the
possibility of constructing a field equation that governs the
gravitational field in the vicinity of this topological dual
matter. We may conclude that both the dynamics of gravitomagnetic
matter and other related topics associated with, for example, the
creation mechanism of gravitomagnetic charge deserve further
consideration.
\\ \\
\textbf{Acknowledgements} This work was supported in part by the
National Natural Science Foundation of China under Project No.
$90101024$.
\\ \\

\textbf{APPENDICES}
\\

\textbf{Appendix 1.} On the Levi-Civita tensor

The Levi-Civita tensor $\varepsilon_{\mu\nu\alpha\beta}$ in the
flat Minkowski spacetime is so defined that it changes sign when
any pair of indices are interchanged and $\varepsilon_{0123}=+1$,
$\varepsilon^{0123}=-1$. The Levi-Civita tensor $\epsilon_{0123}$
and $\epsilon^{0123}$ in the curved spacetime are defined to be
$\epsilon_{0123}=\sqrt{-g}\varepsilon_{0123}$ and
$\epsilon^{0123}=\frac{1}{\sqrt{-g}}\varepsilon^{0123}$,
respectively, where $g$ is the determinant of the matrix
$g_{\mu\nu}$.

One of the most important properties of the Levi-Civita tensor is
that its covariant derivative is vanishing, which may be proved as
follows:

The covariant derivative of $\epsilon_{\mu\nu\theta\tau}$ is given
\begin{equation}
\epsilon_{\mu\nu\theta\tau;\sigma}=\frac{\partial}{\partial
x^{\sigma}}\epsilon_{\mu\nu\theta\tau}-\Gamma^{\lambda}_{\ \
\mu\sigma}\epsilon_{\lambda\nu\theta\tau}-\Gamma^{\lambda}_{\ \
\nu\sigma}\epsilon_{\mu\lambda\theta\tau}-\Gamma^{\lambda}_{\ \
\theta\sigma}\epsilon_{\mu\nu\lambda\tau}-\Gamma^{\lambda}_{\ \
\tau\sigma}\epsilon_{\mu\nu\theta\lambda}.    \eqnum{A1}
\label{appendix11}
\end{equation}
As an illustrative example, we will calculate
$\epsilon_{0123;\sigma}$ only. Here
$\epsilon_{0123}=\sqrt{-g}\varepsilon_{0123}=\sqrt{-g}$. It is
easily obtained that $\frac{\partial}{\partial
x^{\sigma}}\epsilon_{0123}=\frac{1}{2}\sqrt{-g}g^{\alpha\beta}\frac{\partial}{\partial
x^{\sigma}}g_{\alpha\beta}$. If $\mu,\nu,\theta,\tau=0,1,2,3$,
respectively, then by using the antisymmetric property of
$\epsilon_{\mu\nu\theta\tau}$, one can arrive at

\begin{eqnarray}
\Gamma^{\lambda}_{\ \
\mu\sigma}\epsilon_{\lambda\nu\theta\tau}\rightarrow
\Gamma^{\lambda}_{\ \ 0\sigma}\epsilon_{\lambda
123}=\sqrt{-g}\Gamma^{0}_{\ \ 0\sigma},   \qquad
\Gamma^{\lambda}_{\ \
\nu\sigma}\epsilon_{\mu\lambda\theta\tau}\rightarrow
\Gamma^{\lambda}_{\ \
1\sigma}\epsilon_{0\lambda 23}=\sqrt{-g}\Gamma^{1}_{\ \ 1\sigma},           \nonumber \\
\Gamma^{\lambda}_{\ \
\theta\sigma}\epsilon_{\mu\nu\lambda\tau}\rightarrow
\Gamma^{\lambda}_{\ \ 2\sigma}\epsilon_{01\lambda
3}=\sqrt{-g}\Gamma^{2}_{\ \ 2\sigma},       \qquad
\Gamma^{\lambda}_{\ \
\tau\sigma}\epsilon_{\mu\nu\theta\lambda}\rightarrow
\Gamma^{\lambda}_{\ \
3\sigma}\epsilon_{012\lambda}=\sqrt{-g}\Gamma^{3}_{\ \ 3\sigma}.
\eqnum{A2} \label{appendix12}
\end{eqnarray}
So, it follows from (\ref{appendix11}) and (\ref{appendix12}) that
the covariant derivative of the Levi-Civita tensor
$\epsilon_{0123}$ is
\begin{equation}
\epsilon_{0123;\sigma}=\frac{1}{2}\sqrt{-g}g^{\alpha\beta}\frac{\partial}{\partial
x^{\sigma}}g_{\alpha\beta}-\sqrt{-g}\Gamma^{\lambda}_{\ \
\lambda\sigma}\equiv 0.         \eqnum{A3}
\end{equation}

An alternative derivation of the above result is given as follows:

The transformation relation between the Levi-Civita tensors in the
noninertial and local inertial frames of reference satisfies
$\epsilon_{\mu\nu\theta\tau}=\frac{\partial \xi^{a}}{\partial
x^{\mu}} \frac{\partial \xi^{b}}{\partial x^{\nu}} \frac{\partial
\xi^{ac}}{\partial x^{\theta}} \frac{\partial \xi^{a}}{\partial
x^{\tau}} \varepsilon_{abcd}$. So, the derivative of
$\epsilon_{\mu\nu\theta\tau}$ reads

\begin{eqnarray}
\frac{\partial \epsilon_{\mu\nu\theta\tau}}{\partial
x^{\sigma}}&=&\frac{\partial^{2}\xi^{a}}{\partial
x^{\sigma}\partial x^{\mu}}\frac{\partial \xi^{b}}{\partial
x^{\nu}} \frac{\partial \xi^{c}}{\partial x^{\theta}}
\frac{\partial \xi^{d}}{\partial x^{\tau}} \varepsilon_{abcd}
+\frac{\partial \xi^{a}}{\partial x^{\mu}}
\frac{\partial^{2}\xi^{b}}{\partial x^{\sigma}\partial x^{\nu}}
\frac{\partial \xi^{c}}{\partial x^{\theta}}
\frac{\partial \xi^{d}}{\partial x^{\tau}} \varepsilon_{abcd}      \nonumber \\
&+&\frac{\partial \xi^{a}}{\partial x^{\mu}}\frac{\partial
\xi^{b}}{\partial x^{\nu}} \frac{\partial^{2}\xi^{c}}{\partial
x^{\sigma}\partial x^{\theta}} \frac{\partial \xi^{d}}{\partial
x^{\tau}} \varepsilon_{abcd} +\frac{\partial \xi^{a}}{\partial
x^{\mu}}\frac{\partial \xi^{b}}{\partial x^{\nu}} \frac{\partial
\xi^{c}}{\partial x^{\theta}} \frac{\partial^{2}\xi^{d}}{\partial
x^{\sigma}\partial x^{\tau}} \varepsilon_{abcd}.
\eqnum{A4}\label{eqapp}
\end{eqnarray}
The insertion of $\frac{\partial^{2}\xi^{a}}{\partial
x^{\sigma}\partial x^{\mu}}=\Gamma^{\lambda}_{\ \
\sigma\mu}\frac{\partial \xi^{a}}{\partial x^{\lambda}}$,
$\frac{\partial^{2}\xi^{b}}{\partial x^{\sigma}\partial
x^{\nu}}=\Gamma^{\lambda}_{\ \ \sigma\nu}\frac{\partial
\xi^{b}}{\partial x^{\lambda}}$,
$\frac{\partial^{2}\xi^{c}}{\partial x^{\sigma}\partial
x^{\theta}}=\Gamma^{\lambda}_{\ \ \sigma\theta}\frac{\partial
\xi^{c}}{\partial x^{\lambda}}$ and
$\frac{\partial^{2}\xi^{d}}{\partial x^{\sigma}\partial
x^{\tau}}=\Gamma^{\lambda}_{\ \ \sigma\tau}\frac{\partial
\xi^{d}}{\partial x^{\lambda}}$ into (\ref{eqapp}) yields

\begin{equation}
\frac{\partial \epsilon_{\mu\nu\theta\tau}}{\partial
x^{\sigma}}=\Gamma^{\lambda}_{\ \
\sigma\mu}\epsilon_{\lambda\nu\theta\tau}+\Gamma^{\lambda}_{\ \
\sigma\nu}\epsilon_{\mu\lambda\theta\tau}+\Gamma^{\lambda}_{\ \
\sigma\theta}\epsilon_{\mu\nu\lambda\tau}+\Gamma^{\lambda}_{\ \
\sigma\tau}\epsilon_{\mu\nu\theta\lambda}.
\eqnum{A5}\label{eqa}
\end{equation}

It follows from Eq.(\ref{appendix11}) and (\ref{eqa}) that
$\epsilon_{\mu\nu\theta\tau;\sigma}=0$. Thus we have proven that
the covariant derivative of the Levi-Civita tensor truly vanishes.

In the four-dimensional spacetime, the following four identities
(which are the inner products of two Levi-Civita tensors) are
useful for investigating the dynamics of gravitomagnetic matter:
 \begin{eqnarray}
\epsilon_{\mu\nu\alpha\beta}\epsilon^{\mu\nu\alpha\beta}&=&-4!,                 \nonumber \\
\epsilon_{\mu\nu\alpha\beta}\epsilon^{\lambda\nu\alpha\beta}&=&-3!g^{\lambda}_{\mu},
\nonumber \\
\epsilon_{\mu\nu\alpha\beta}\epsilon^{\lambda\sigma\alpha\beta}&=&-2!\left(g^{\lambda}_{\mu}g^{\sigma}_{\nu}-g^{\sigma}_{\mu}g^{\lambda}_{\nu}\right),
\nonumber \\
\epsilon_{\mu\nu\alpha\beta}\epsilon^{\lambda\sigma\tau\beta}&=&-\left(g^{\lambda}_{\mu}g^{\sigma}_{\nu}g^{\tau}_{\alpha}+g^{\tau}_{\mu}g^{\lambda}_{\nu}g^{\sigma}_{\alpha}+g^{\sigma}_{\mu}g^{\tau}_{\nu}g^{\lambda}_{\alpha}-g^{\sigma}_{\mu}g^{\lambda}_{\nu}g^{\tau}_{\alpha}-g^{\tau}_{\mu}g^{\sigma}_{\nu}g^{\lambda}_{\alpha}-g^{\lambda}_{\mu}g^{\tau}_{\nu}g^{\sigma}_{\alpha}\right).
\eqnum{A6}\label{eq71}
\end{eqnarray}
\\

\textbf{Appendix 2.} Proving the first identity of Eq.(\ref{eq34})

According to the fourth identity of Eq.(\ref{eq71}), one can
arrive at
\begin{eqnarray}
\epsilon^{\theta\tau\omega\nu}\epsilon^{\mu}_{\ \ \tau}\
^{\lambda\sigma}&=&\epsilon^{\tau\theta\omega\nu}\epsilon_{\tau}^{\
\ \mu\lambda\sigma}                  \nonumber \\
&=&-\left(g^{\mu\theta}g^{\omega\lambda}g^{\nu\sigma}+g^{\nu\mu}g^{\theta\lambda}g^{\omega\sigma}+g^{\omega\mu}g^{\nu\lambda}g^{\theta\sigma}-g^{\mu\nu}g^{\omega\lambda}g^{\theta\sigma}-g^{\mu\theta}g^{\nu\lambda}g^{\omega\sigma}-g^{\omega\mu}g^{\theta\lambda}g^{\nu\sigma}\right).
\eqnum{A7}
\end{eqnarray}
So, the further calculation yields
\begin{eqnarray}
\epsilon^{\theta\tau\omega\nu}\tilde{{\mathcal R}}^{\mu}_{\ \
\tau\omega\nu}&=&\frac{1}{2}\epsilon^{\tau\theta\omega\nu}\epsilon_{\tau}^{\
\ \mu\lambda\sigma}{\mathcal R}_{\lambda\sigma\omega\nu}                      \nonumber \\
&=&-\frac{1}{2}\left(g^{\mu\theta}{\mathcal R}^{\omega\nu}_{\ \ \
\omega\nu}+{\mathcal R}^{\theta\omega}_{\ \ \ \omega}\
^{\mu}+{\mathcal R}^{\nu\theta\mu}_{\ \ \ \ \nu}-{\mathcal
R}^{\omega\theta}_{\ \ \ \omega}\ ^{\mu}-g^{\mu\theta}{\mathcal
R}^{\nu\omega}_{\ \ \ \omega\nu}-{\mathcal R}^{\theta\nu\mu}_{\ \
\ \ \nu}\right)                                \nonumber \\
&=&-\left({\mathcal
R}^{\theta\mu}-\frac{1}{2}g^{\mu\theta}{\mathcal
R}\right)-\left({\mathcal
R'}^{\theta\mu}-\frac{1}{2}g^{\mu\theta}{\mathcal R}\right).
\eqnum{A8}
\end{eqnarray}
If the Ricci tensor ${\mathcal R}^{\theta\mu}$ and ${\mathcal
R'}^{\theta\mu}$ are respectively defined to be ${\mathcal
R}^{\theta\mu}={\mathcal R}^{\nu\theta\mu}_{\ \ \ \ \nu}$,
${\mathcal R'}^{\theta\mu}=g_{\lambda\tau}{\mathcal
R}^{\theta\lambda\tau\mu}$, and the curvature scalar ${\mathcal
R}$ is obtained via ${\mathcal R}^{\omega\nu}_{\ \ \
\omega\nu}=-g_{\alpha\beta}{\mathcal R}^{\alpha\beta}=-{\mathcal
R}$, then we can readily show that this identity
\begin{equation}
\frac{1}{2}\epsilon^{\theta\tau\omega\nu}\tilde{{\mathcal
R}}^{\mu}_{\ \
\tau\omega\nu}=-\frac{1}{2}\left(G^{\theta\mu}+G'^{\theta\mu}\right)
\eqnum{A9}
\end{equation}
truly holds.
\\

\textbf{Appendix 3.}
 Showing that $\frac{1}{2}\int_{\Omega}\sqrt{-g}\epsilon^{\mu\tau\omega\nu}\delta{\mathcal R}_{\tau\omega\nu\mu}{\rm d}\Omega$ is a surface
 integral

For the first we obtain the variation of ${\mathcal
R}_{\tau\omega\nu\mu}$
 \begin{eqnarray}
\delta{\mathcal R}_{\tau\omega\nu\mu}&=&\frac{\partial}{\partial
x^{\nu}}\delta \Gamma_{\tau, \omega\mu}-\frac{\partial}{\partial
x^{\mu}}\delta \Gamma_{\tau, \omega\nu}-\delta\Gamma^{\lambda}_{\
\ \omega\mu }\Gamma_{\lambda,\tau\nu}-\Gamma^{\lambda}_{\ \
\omega\mu
}\delta\Gamma_{\lambda,\tau\nu}+\delta\Gamma^{\lambda}_{\ \
\omega\nu }\Gamma_{\lambda,\tau\mu}+\Gamma^{\lambda}_{\ \
\omega\nu }\delta\Gamma_{\lambda,\tau\mu}                  \nonumber \\
&=&\left(\frac{\partial}{\partial x^{\nu}}\delta \Gamma_{\tau,
\omega\mu}-\Gamma^{\lambda}_{\ \
\tau\nu}\delta\Gamma_{\lambda,\omega\mu}-\Gamma^{\lambda}_{\ \
\mu\nu
}\delta\Gamma_{\tau,\omega\lambda}\right)-\left(\frac{\partial}{\partial
x^{\mu}}\delta \Gamma_{\tau, \omega\nu}-\Gamma^{\lambda}_{\ \
\tau\mu}\delta\Gamma_{\lambda,\omega\nu}-\Gamma^{\lambda}_{\ \
\nu\mu }\delta\Gamma_{\tau,\omega\lambda}\right)
\nonumber \\
&+&\Gamma^{\lambda}_{\ \ \omega\nu
}\delta\Gamma_{\lambda,\tau\mu}-\Gamma^{\lambda}_{\ \ \omega\mu
}\delta\Gamma_{\lambda,\tau\nu}.      \eqnum{A10} \label{eq75}
\end{eqnarray}
For the second, we analyze the third term $\Gamma^{\lambda}_{\ \
\omega\nu }\delta\Gamma_{\lambda,\tau\mu}$ on the right-handed
side of Eq.(\ref{eq75}). The following calculation is trivial
\begin{eqnarray}
\epsilon^{\mu\tau\omega\nu}\Gamma^{\lambda}_{\ \ \omega\nu
}\delta\Gamma_{\lambda,\tau\mu}&=&\frac{1}{2}\epsilon^{\mu\tau\omega\nu}\Gamma^{\lambda}_{\
\ \omega\nu
}\delta\left(\frac{\partial g_{\mu\lambda}}{\partial x^{\tau}}-\frac{\partial g_{\tau\mu}}{\partial x^{\lambda}}\right)                  \nonumber \\
&=&-\epsilon^{\mu\tau\omega\nu}\Gamma^{\lambda}_{\ \ \omega\nu
}\delta \frac{1}{2}\left(-\frac{\partial g_{\mu\lambda}}{\partial
x^{\tau}}+\frac{\partial g_{\tau\mu}}{\partial
x^{\lambda}}+\frac{\partial g_{\lambda\tau}}{\partial
x^{\mu}}\right)=-\epsilon^{\mu\tau\omega\nu}\Gamma^{\lambda}_{\ \
\omega\nu }\delta\Gamma_{\tau,\lambda\mu}.      \eqnum{A11}
\label{eq76}
\end{eqnarray}
In the same manner, we can rewrite
$-\epsilon^{\mu\tau\omega\nu}\Gamma^{\lambda}_{\ \ \omega\mu
}\delta\Gamma_{\lambda,\tau\nu}$ as follows
\begin{equation}
-\epsilon^{\mu\tau\omega\nu}\Gamma^{\lambda}_{\ \ \omega\mu
}\delta\Gamma_{\lambda,\tau\nu}=\epsilon^{\mu\tau\omega\nu}\Gamma^{\lambda}_{\
\ \omega\mu }\delta\Gamma_{\tau,\lambda\nu}.
\eqnum{A12}\label{eq77}
\end{equation}
So, it follows from both Eq.(\ref{eq76}) and (\ref{eq77}) that one
can obtain
\begin{equation}
\epsilon^{\mu\tau\omega\nu}\left(\Gamma^{\lambda}_{\ \ \omega\nu
}\delta\Gamma_{\lambda,\tau\mu}-\Gamma^{\lambda}_{\ \ \omega\mu
}\delta\Gamma_{\lambda,\tau\nu}\right)=\epsilon^{\mu\tau\omega\nu}\left(-\Gamma^{\lambda}_{\
\ \omega\nu }\delta\Gamma_{\tau,\lambda\mu}+\Gamma^{\lambda}_{\ \
\omega\mu }\delta\Gamma_{\tau,\lambda\nu}\right),   \eqnum{A13}
\end{equation}
which will lead to some changes in the expression for
$\epsilon^{\mu\tau\omega\nu}\delta{\mathcal
R}_{\tau\omega\nu\mu}$. Thus the integrand
$\epsilon^{\mu\tau\omega\nu}\delta{\mathcal R}_{\tau\omega\nu\mu}$
may be rewritten as
\begin{eqnarray}
\epsilon^{\mu\tau\omega\nu}\delta{\mathcal
R}_{\tau\omega\nu\mu}&=&\epsilon^{\mu\tau\omega\nu}\left(\frac{\partial}{\partial
x^{\nu}}\delta \Gamma_{\tau, \omega\mu}-\Gamma^{\lambda}_{\ \
\tau\nu}\delta\Gamma_{\lambda,\omega\mu}-\Gamma^{\lambda}_{\ \
\omega\nu }\delta\Gamma_{\tau,\lambda\mu}-\Gamma^{\lambda}_{\ \
\mu\nu
}\delta\Gamma_{\tau,\omega\lambda}\right)                 \nonumber \\
&-&\epsilon^{\mu\tau\omega\nu}\left(\frac{\partial}{\partial
x^{\mu}}\delta \Gamma_{\tau, \omega\nu}-\Gamma^{\lambda}_{\ \
\tau\mu}\delta\Gamma_{\lambda,\omega\nu}-\Gamma^{\lambda}_{\ \
\omega\mu }\delta\Gamma_{\tau,\lambda\nu}-\Gamma^{\lambda}_{\ \
\nu\mu }\delta\Gamma_{\tau,\omega\lambda}\right)
 \nonumber \\
&=&\epsilon^{\mu\tau\omega\nu}\left[\left(\delta\Gamma_{\tau,\omega\mu}\right)_{;\nu}-\left(\delta\Gamma_{\tau,\omega\nu}\right)_{;\mu}\right].
\eqnum{A14}
\end{eqnarray}
By using
 $\epsilon^{\mu\tau\omega\nu}_{;\nu}=\epsilon^{\mu\tau\omega\nu}_{;\mu}=0$,
 one can therefore arrive at
\begin{equation}
\frac{1}{2}\epsilon^{\mu\tau\omega\nu}\delta{\mathcal
R}_{\tau\omega\nu\mu}=B^{\mu}_{\
;\mu}=\frac{1}{\sqrt{-g}}\frac{\partial}{\partial
x^{\mu}}\left(\sqrt{-g}B^{\mu}\right) \quad {\rm with} \quad
B^{\mu}=-\epsilon^{\mu\tau\omega\nu}\delta\Gamma_{\tau,\omega\nu}.
\eqnum{A15}
\end{equation}
Hence, we have shown that
$\frac{1}{2}\int_{\Omega}\sqrt{-g}\epsilon^{\mu\tau\omega\nu}\delta{\mathcal
R}_{\tau\omega\nu\mu}{\rm d}\Omega$ is truly a surface integral,
{\it i.e.},
\begin{equation}
\frac{1}{2}\int_{\Omega}\sqrt{-g}\epsilon^{\mu\tau\omega\nu}\delta{\mathcal
R}_{\tau\omega\nu\mu}{\rm
d}\Omega=\int_{\Omega}\frac{\partial}{\partial
x^{\mu}}\left(\sqrt{-g}B^{\mu}\right){\rm d}\Omega,
\eqnum{A16}
\end{equation}
which is a four-dimensional volume integral of the divergence
$\frac{\partial}{\partial x^{\mu}}\left(\sqrt{-g}B^{\mu}\right)$,
and hence has no contribution to the derivation of the dual
Einstein tensor.
\\

\textbf{Appendix 4.} The variation of the Levi-Civita tensor
$\epsilon_{\mu}^{\ \ \lambda\sigma\tau}$

With the help of the two expressions $\delta
\frac{1}{\sqrt{-g}}=\frac{1}{2}\frac{1}{\sqrt{-g}}g_{\alpha\beta}\delta
g^{\alpha\beta} $ and $\delta
g_{\mu\theta}=-g_{\mu\alpha}g_{\theta\beta}\delta
g^{\alpha\beta}$, one can reach

\begin{eqnarray}
\delta \epsilon_{\mu}^{\ \ \lambda\sigma\tau}=\delta
\left(g_{\mu\theta}\epsilon^{\theta\lambda\sigma\tau}\right)&=&\delta
g_{\mu\theta}
\epsilon^{\theta\lambda\sigma\tau}+g_{\mu\theta}\delta\epsilon^{\theta\lambda\sigma\tau}                   \nonumber \\
&=&\left(-g_{\mu\alpha}g_{\theta\beta}\epsilon^{\theta\lambda\sigma\tau}+\frac{1}{2}g_{\mu\theta}g_{\alpha\beta}\epsilon^{\theta\lambda\sigma\tau}\right)\delta
g^{\alpha\beta}.         \eqnum{A17}
\end{eqnarray}
Thus we obtain the variation of the Levi-Civita tensor
$\epsilon_{\mu}^{\ \ \lambda\sigma\tau}$.
\\

\textbf{Appendix 5.} Mid-dual and two-side dual curvature tensors

In Sec. III, we only considered the left-dual and right-dual
Riemann (and hence Ricci) curvature tensors, and showed that
$\frac{1}{2}\epsilon_{\nu}^{\ \ \tau\lambda\sigma}{\mathcal
R}_{\mu\tau\lambda\sigma}\equiv 0$, $\frac{1}{2}\epsilon_{\nu}^{\
\ \mu\tau\lambda}{\mathcal
R}_{\mu\tau\lambda\sigma}=\tilde{{\mathcal R}}_{\nu\sigma}$. In
fact, there exist other kinds of dual Riemann tensors. See, for
example, $\frac{1}{2}\epsilon_{\nu}^{\ \
\mu\lambda\sigma}{\mathcal R}_{\mu\tau\lambda\sigma}$. It follows
from the expression ${\mathcal
R}_{\mu\tau\lambda\sigma}=\frac{\partial}{\partial
x^{\lambda}}\Gamma_{\mu,\tau\sigma}-\frac{\partial}{\partial
x^{\sigma}}\Gamma_{\mu,\tau\lambda}-\Gamma^{\theta}_{\ \
\tau\sigma}\Gamma_{\theta,\mu\lambda}+\Gamma^{\theta}_{\ \
\tau\lambda}\Gamma_{\theta,\mu\sigma}$ that
\begin{equation}
{\mathcal R}_{\mu\tau\lambda\sigma}=-{\mathcal
R}_{\tau\mu\lambda\sigma}+\left(\partial_{\lambda}\partial_{\sigma}-\partial_{\sigma}\partial_{\lambda}\right)g_{\mu\tau}.
\eqnum{A18}
\end{equation}
Thus one can obtain
\begin{equation}
\frac{1}{2}\epsilon_{\nu}^{\ \ \mu\lambda\sigma}{\mathcal
R}_{\mu\tau\lambda\sigma}=\frac{1}{2}\epsilon_{\nu}^{\ \
\mu\lambda\sigma}\left(\partial_{\lambda}\partial_{\sigma}-\partial_{\sigma}\partial_{\lambda}\right)g_{\mu\tau}.
\eqnum{A19}
\end{equation}
Note that here $\frac{1}{2}\epsilon_{\nu}^{\ \
\mu\lambda\sigma}\left(\partial_{\lambda}\partial_{\sigma}-\partial_{\sigma}\partial_{\lambda}\right)g_{\mu\tau}=\epsilon_{\nu}^{\
\
\mu\lambda\sigma}\partial_{\lambda}\partial_{\sigma}g_{\mu\tau}=\frac{1}{2}\epsilon_{\nu}^{\
\
\mu\lambda\sigma}\partial_{\lambda}\left(\partial_{\sigma}g_{\mu\tau}-\partial_{\mu}g_{\sigma\tau}\right)=-2\tilde{{\mathcal
R}}_{\nu\tau}$. So, a useful fact that no essential difference
exists between $\frac{1}{2}\epsilon_{\nu}^{\ \
\mu\lambda\sigma}{\mathcal R}_{\mu\tau\lambda\sigma}$ and
$\tilde{{\mathcal R}}_{\nu\tau}$ is thus demonstrated. This,
therefore, means that we need not take account of
$\frac{1}{2}\epsilon_{\nu}^{\ \ \mu\lambda\sigma}{\mathcal
R}_{\mu\tau\lambda\sigma}$ in considering the dynamics of
gravitomagnetic charge.

In what follows, we will discuss the mid-dual and two-side dual
curvature tensors. It will be shown that these two kinds of dual
curvature tensors are either trivial or equivalent to the
left-dual tensor.

The mid-dual Riemann tensor under consideration is
$\frac{1}{2}\epsilon_{\tau\omega}^{\ \ \ \lambda\sigma}{\mathcal
R}_{\mu\lambda\sigma\nu}$, and the corresponding dual Ricci
tensors can be written as
\begin{equation}
g^{\tau\omega}\left(\frac{1}{2}\epsilon_{\tau\omega}^{\ \ \
\lambda\sigma}{\mathcal R}_{\mu\lambda\sigma\nu}\right)=0,   \quad
g^{\omega\nu}\left(\frac{1}{2}\epsilon_{\tau\omega}^{\ \ \
\lambda\sigma}{\mathcal R}_{\mu\lambda\sigma\nu}\right)=0,   \quad
g^{\omega\mu}\left(\frac{1}{2}\epsilon_{\tau\omega}^{\ \ \
\lambda\sigma}{\mathcal
R}_{\mu\lambda\sigma\nu}\right)=\tilde{{\mathcal R}}_{\tau\nu},
...       \eqnum{A20} \label{eqmid}
\end{equation}
It is readily verified that the mid-dual tensors in (\ref{eqmid})
are either trivial or equivalent to the left-dual tensor.

The two-side dual Riemann curvature tensor is defined to be
$\frac{1}{2}\epsilon_{\mu\nu}^{\ \ \ \lambda\sigma}{\mathcal
R}_{\lambda\tau\omega\sigma}$, the corresponding Ricci tensors of
which are given as follows
\begin{equation}
g^{\mu\nu}\left(\frac{1}{2}\epsilon_{\mu\nu}^{\ \ \
\lambda\sigma}{\mathcal R}_{\lambda\tau\omega\sigma}\right)=0,
\quad      g^{\mu\tau}\left(\frac{1}{2}\epsilon_{\mu\nu}^{\ \ \
\lambda\sigma}{\mathcal
R}_{\lambda\tau\omega\sigma}\right)=-\tilde{{\mathcal
R}}_{\nu\mu},    \quad
g^{\mu\omega}\left(\frac{1}{2}\epsilon_{\mu\nu}^{\ \ \
\lambda\sigma}{\mathcal
R}_{\lambda\tau\omega\sigma}\right)=-\frac{1}{2}\tilde{{\mathcal
R}}_{\nu\tau}, ...        \eqnum{A21}
\end{equation}
which are also trivial or equivalent to the left-dual tensor.

For this reason, in this paper we will not consider further other
kinds of dual tensors such as mid-dual and two-side dual curvature
tensors.
\\

\textbf{Appendix 6.} The duality relation between the
electromagnetic field equations of magnetic and electric charges

The electromagnetic field equations of magnetic and electric
charges are of the form (where $j^{\nu}_{\rm e}$ and $j^{\nu}_{\rm
m}$ are the electric and magnetic charge current densities,
respectively)
\begin{equation}
\partial_{\mu}{\mathcal F}^{\mu\nu}=j^{\nu}_{\rm e},   \qquad       \partial_{\mu}\tilde{{\mathcal F}}^{\mu\nu}=j^{\nu}_{\rm m}
\eqnum{A22}
\end{equation}
with ${\mathcal
F}_{\alpha\beta}=\partial_{\alpha}A_{\beta}-\partial_{\beta}A_{\alpha}$,
$\tilde{{\mathcal
F}}^{\mu\nu}=\frac{1}{2}\epsilon^{\mu\nu\alpha\beta}{\mathcal
F}_{\alpha\beta}$,   ${\mathcal
F}^{\mu\nu}=-\frac{1}{2}\epsilon^{\mu\nu\alpha\beta}\tilde{{\mathcal
F}}_{\alpha\beta}$. The above electrodynamics is developed in the
framework of the electromagnetic field $A_{\mu}$. In contrast, we
can also propose a dual theory to the above one, where the
formalism is constructed in the framework of the so-called dual
electromagnetic field $\tilde{A}_{\mu}$, {\it i.e.},
\begin{equation}
\tilde{{\mathcal
F}}^{\mu\nu}=\partial^{\mu}\tilde{A}^{\nu}-\partial^{\nu}\tilde{A}^{\mu},
\qquad    {\mathcal
F}^{\mu\nu}=-\frac{1}{2}\epsilon^{\mu\nu\alpha\beta}\tilde{{\mathcal
F}}_{\alpha\beta}.
\eqnum{A23}
\end{equation}
The above theories are dual to each other and, furthermore,
equivalent to one another. This, therefore, implies that one can
study the electrodynamics not only in the framework of $A_{\mu}$
but also in the one of $\tilde{A}_{\mu}$. In Sec. VIII, the
similar dual phenomenon in gravity theory is discussed, where a
second-rank antisymmetric tensor $h_{\mu\nu}$ is defined to
describe the gravitational field. So Einstein's field equation can
also be expressed inside the framework of $h_{\mu\nu}$, just as
the fact that the electrodynamics can be described within the
framework of $\tilde{A}_{\mu}$.


\begin{references}
\bibitem{Shen} Shen, J.Q. (2002). {\it Gen. Relativ. Gravit.} {\bf
34}, 1423 [See also in {\it arXiv: gr-qc/0301067}].

\bibitem{Shenarxiv}  Shen, J.Q. (2003). {\it arXiv:
gr-qc/0301100}.

\bibitem{Zimmerman} Zimmerman, R.L. and Shahir, B.Y. (1989). {\it Gen. Relativ. Gravit.} {\bf
21}, 821.

\bibitem{Newman} Newman, E.T., Tamburino, and Unti, T. (1963). {\it J. Math.
Phys.} {\bf 4}, 915.

\bibitem{Demiansky} Demiansky, M. and Newman, E.T. (1966). {\it Bull. Acad. Pol. Sci. Ser. Math. Astron.
Phys.} {\bf XIV}, 653.

\bibitem{Dowker} Dowker, J.S. and Roche, J.A. (1967). {\it Proc. Phys. Soc.
London} {\bf 92}, 1.

\bibitem{Nouri-Zonoz} Nouri-Zonoz, M. (1997). {\it ArXiv: gr-qc/9706015}.

\bibitem{Lynden-Bell}  Lynden-Bell, D. and Nouri-Zonoz, M. (1998).
{\it Rev. Mod. Phys.} {\bf 70}, 427.

\bibitem{Miller} Miller, J.G. (1973). {\it J. Math. Phys.} {\bf
14}, 486.

\bibitem{Bini1} Bini, D., Cherubini, C., Jantzen, R.T. and
Mashhoon, B. (2003). {\it ArXiv: gr-qc/0301080}.

\bibitem{Bini2}  Bini, D., Cherubini, and Jantzen, R.T. (2002). {\it ArXiv: gr-qc/0210003}.


\bibitem{Dowker2}  Dowker, J.S. (1974). {\it Gen. Relativ. Gravit.} {\bf
5}, 603.

\bibitem{Zee}  Zee, A. (1985). {\it Phys. Rev. Lett.} {\bf 55},
2379.

\bibitem{Zeleny}  Zeleny, W.B. (1991). {\it Am. J. Phys.} {\bf 59},
412.

\bibitem{Wu} Wu, T.T. and Yang, C.N. (1969). {\it Properties of matter under unusual
conditions} (eds. Mark, H. and Fernback, P.).

\bibitem{Ni} Hehl, F.W. and Ni, W.T. (1990). {\it Phys. Rev.} D {\bf
42}, 2045.

\bibitem{Mashhoon1} Mashhoon, B. (2000). {\it Class. Quant. Gravit.} {\bf 17}, 2399.

\bibitem{Mashhoon2}  Mashhoon, B. (1999). {\it Gen. Relativ. Gravit.} {\bf 31}, 681.

\bibitem{Gao} Gao, X.C., Fu, J. and Shen, J.Q. (2000). {\it Eur. Phys.
J.} C {\bf 13}, 527.

\bibitem{Overduin} Overduin, J.M. and Wesson, P.S. (1997). {\it Phys.
Rep.} {\bf 283}, 303.

\bibitem{Weinberg} Weinberg, S. (1989). {\it Rev. Mod.
Phys.} {\bf 61}, 1.

\bibitem{Datta} Datta, D.P. (1995). {\it  Gen. Relativ. Gravit.} {\bf 27},
341.

\bibitem{Krauss}  Krauss, L.M. and Turner, M. (1995).
{\it Gen. Relativ. Gravit.} {\bf 27}, 1137.

\bibitem{Kakushadze}  Kakushadze, Z. (2000). {\it Phys. Lett.} B {\bf 488}, 402.

\bibitem{Hooft}  Hooft, G 't. (1974). {\it Nucl. Phys.} B {\bf 79}, 276.

\end{references}
\end{document}